\documentclass{aa}
\usepackage[english]{babel}
\usepackage{microtype}
\usepackage{txfonts}

\usepackage{mathtools}
\usepackage{bm}
\usepackage{mathrsfs}
\usepackage{nccmath}

\usepackage[debug]{hyperref}
\hypersetup{colorlinks=true, allcolors=blue}

\usepackage{graphicx}

\makeatletter
\DeclareRobustCommand\onedot{\futurelet\@let@token\@onedot}
\def\@onedot{\ifx\@let@token.\else.\null\fi\xspace}

\makeatother

\renewcommand{\vec}[1]{\bm{#1}}

\newcommand{\grad}{\vec{\nabla}}
\newcommand{\dt}{\partial_t}
\newcommand{\Dt}{{\mathscr{D}_t}}
\renewcommand{\dot}[1] {\overset{\,_{\mbox{\Large .}}}{#1}}

\newcommand{\DTad}{{\nabla_{\mathrm{ad}}}}

\begin{document}

\title{
    Phase-space averaging for stellar convection
}
\subtitle{
        II. Maximum-entropy closures for mixed radiative--convective envelopes
}

\author{
        P.~S.~Houdayer \email{pierre.houdayer@utoulouse.fr}
        \and
        M.~Rieutord \email{michel.rieutord@utoulouse.fr}
}

\institute{
        Univ Toulouse, CNRS, CNES, IRAP, 14, avenue Édouard Belin, F-31400 Toulouse, France.
}

\date{Received 28 May 2026; Accepted 10 September 2026}

\abstract
{Surface convection in cool stars sets the entropy jump between the atmosphere and the deep convective envelope and therefore affects the radius and structure of one-dimensional stellar models.
In standard local treatments, this entropy jump is controlled through an effective convective efficiency.}
{We have developed a phase-space closure that relates the mean stratification to the local distribution of entropy and velocity among convective elements.
Our aim is to describe how the surface entropy transition emerges from the statistics of these elements rather than prescribing it directly.}
{We first constructed a maximum-entropy model for the convective bulk, assuming that the realised states are bounded by a deep entropy level \(S\).
This prediction was then confronted with three-dimensional radiation-hydrodynamics simulations from the M3DIS grid.
The comparison motivated an extension to a two-population closure in which an optimal bulk component coexists with a radiatively cooled plume component.}
{The maximum-entropy construction predicts a one-sided exponential entropy distribution, and its low-entropy tail is recovered in the deep convective part of the simulations.
Closer to the surface, the distribution is reorganised by radiative cooling and separates into two thermodynamic components.
Using a radiative entropy branch, an adiabatically descending cooled branch, and a population fraction controlled by optical depth, the two-population closure reproduces the mean entropy stratification of the four reference simulations over most of the displayed range without adjusting the entropy profile itself.}
{The surface transition can therefore be interpreted as a continuous population decomposition, rather than as a sharp boundary between radiative and convective layers.
This provides a natural route towards one-dimensional stellar models in which the surface entropy jump is determined from the statistical organisation of the flow instead of being set by a calibrated mixing-length efficiency.}


\maketitle
\nolinenumbers

\section{Introduction}
\label{sec:introduction}

In stellar envelopes, thermal convection develops where radiative diffusion can no longer carry the imposed energy flux without destabilising the structure.
This situation commonly occurs in layers where the opacity rises sharply, in particular in partial-ionisation regions and more generally throughout the outer envelope where the thermodynamic state changes rapidly with depth.
As the radiative conductivity decreases, the temperature gradient required to transport the stellar flux by radiation steepens accordingly.
Once this gradient exceeds the stability threshold of the entropy stratification, buoyancy-driven motions appear and begin to contribute to the flux transport.
The emergence of thermal convection is therefore tied to a redistribution of energy transport between radiative diffusion and fluid motions, whose respective contributions vary continuously across the envelope.

This qualitative picture, however, still leaves open a central question for stellar modelling.
Once convection has appeared, its transport feeds back on the very structure that made it necessary in the first place.
By modifying the temperature profile, it alters the local opacity, the radiative contribution to the flux, and thus the amount of energy that must still be carried by fluid motions.
The problem is therefore not only to determine where convection is triggered but also which mean state emerges from this radiative--convective feedback.
The relevant questions are how efficient convection becomes in the deep envelope, where the stratification is nearly adiabatic; how this efficiency changes as the surface is approached and radiative exchange becomes increasingly important on dynamical timescales; and how the transition is organised between the deep convective layers, the radiatively controlled surface region, and the surrounding stable stratification.
These are precisely the questions that one-dimensional convection models are meant to address when they are used to predict the mean structure of a star.

In practice, the structure of a stellar envelope is governed by the mean equations of hydrostatic balance and energy transport.
Once hydrostatic equilibrium has been imposed, the remaining difficulty is essentially thermal: one must determine the thermal stratification that allows the stellar flux to be carried through the envelope.
In one-dimensional models, however, the mechanisms responsible for this transport cannot be followed in their full dynamical complexity \citep{Kupka2017,Joyce2023}.
Their influence on the mean structure must instead be represented through simplified closures, typically through phenomenological prescriptions based on mixing-length theory (MLT) or related formulations \citep{BohmVitense1958,Spiegel1963,Gough1977,Canuto1991,Joyce2023}.
The focus of the problem then shifts somewhat.
The global issues raised above---how the mean state is selected, how efficient convection becomes, and how the transition proceeds near the surface---must be recast into local prescriptions for the transport processes themselves.
One must specify, for example, how rapidly fluid elements exchange heat with their surroundings, how far they travel before losing their identity, and how efficiently their motions contribute to the mean flux \citep{BohmVitense1958,Gough1977}.
It is this set of local ingredients, rather than the explicit flow dynamics, that closes the one-dimensional problem and ultimately determines the mean thermal stratification predicted by the model.

Among the variables used to describe the resulting structure, entropy is especially convenient.
In the deep parts of convective envelopes, transport is sufficiently efficient for the stratification to remain close to adiabatic so that the entropy varies only weakly with depth and the envelope is characterised by an almost constant deep-entropy level.
Towards the surface, however, the situation changes qualitatively.
As the density drops, the ability of fluid motions to carry a given flux decreases accordingly, while the lower sound speed limits the velocities that can be reached without entering a strongly compressible regime \citep{Stein1989a,Cattaneo1990,Cattaneo1991,Nordlund2009}.
At the same time, radiative losses become increasingly effective, and thermal stratification departs from adiabaticity \citep{Stein1998,Nordlund2009,Magic2016}.
To a large extent, the thermal structure predicted by a one-dimensional model may therefore be viewed as the combination of two simple features: a deep convective adiabat and a near-surface entropy jump connecting it to the radiatively controlled atmosphere \citep{Trampedach2014,Magic2016,Tanner2016,Sonoi2019}.
In that sense, the efficiency of convection is reflected directly in the entropy profile itself through both the level of the deep adiabat and the depth, width, and amplitude of the entropy drop that develops across the surface boundary layers.

The aim of the present paper is to construct one-dimensional thermal closures within the phase-space averaging (PSA) framework introduced in \citep[][hereafter Paper~I]{Houdayer2026a}.
In that framework, the mean equations are obtained as velocity-space moments of an underlying phase-space mass distribution so that closing the one-dimensional equations amounts to specifying that distribution under a limited set of mean constraints.
In the convective bulk, where the realised part of the local phase-space flow is predominantly expansive, the associated production of statistical entropy provides a physical motivation for a maximum-entropy closure of the local distribution compatible with those constraints.

We begin by examining the mean envelope structure predicted by this closure.
This construction clarifies which aspects of the envelope can already be captured by a reduced thermodynamic description, but it also shows that a single-population closure cannot, in general, reconstruct the full mean entropy profile, especially across the surface transition where distinct thermal populations coexist.
Guided by three-dimensional simulations, we then used these limitations to motivate a multi-population extension aimed at reconstructing the mean entropy profile, in particular the surface entropy transition, while remaining within the same one-dimensional framework.

The paper is organised as follows.
In Sect.~\ref{sec:1D_closure}, we derive a one-dimensional reduction of the averaged equations and construct an entropy-maximising closure for the fluid-particle distribution, which is used as a reference description of the convective envelope.
In Sect.~\ref{sec:phenomenology_3D}, we then focus on three-dimensional envelope simulations to identify the thermodynamic partition of the phase-space distribution that must be retained in a model of the surface transition.
Building on this information, in Sect.~\ref{sec:multi_population}, we introduce the general multi-population formalism and develop its simplest practical realisation, namely a two-population closure for the surface transition.
The resulting model is then confronted with three-dimensional simulations in Sect.~\ref{sec:comparison_3D}.
Finally, the assumptions, limitations, and implications of the framework are discussed in Sect.~\ref{sec:discussion}, and the main conclusions are summarised in Sect.~\ref{sec:conclusion}.

\section{One-dimensional closure for the convective bulk}
\label{sec:1D_closure}

\subsection{Phase-space framework and averaged equations}
\label{sec:psa_mean_equations}

We briefly recall the elements of the phase-space averaging framework needed in the present paper.
In this approach, the flow is described in terms of mesoscopic fluid particles distributed in phase space, whose mass distribution is represented by the phase-space mass density \(\rho(t,\vec r,\vec u)\), with \(\vec r\) the position and \(\vec u\) the velocity, so that
\begin{equation}
        dM = \rho(t,\vec r,\vec u) \, d\vec r \, d\vec u
\end{equation}
is the mass of fluid particles located near \(\vec r\) with velocity near \(\vec u\) at time \(t\).
The evolution of \(\rho\) is governed by the phase-space continuity equation
\begin{equation}
        \label{eq:phase_space_continuity}
        \partial_t \rho + \grad_{r} \cdot(\rho \vec u) + \grad_{u} \cdot(\rho \vec\gamma) = 0,
\end{equation}
where \(\vec\gamma\) denotes the acceleration of a fluid particle arising from local hydrodynamic forces.

Under the assumptions of a high Reynolds number and immediate pressure adjustment, as discussed in Paper~I, the same phase-space dynamics can be written in Liouvillian form through a local generator \(d\psi\) defined by
\begin{equation}
        d\psi = dH - T ds,
        \qquad
        H = \frac{\vec u^{\,2}}{2} + \varepsilon + pv + \Phi,
\end{equation}
where \(H\) is the fluid-particle Hamiltonian, collecting the kinetic energy, the internal energy \(\varepsilon\), and the thermodynamic and gravitational potentials \(pv\) and \(\Phi\), while \(s\) and \(T\) denote the specific entropy and temperature.
In this formulation, the fluid-particle motion is generated in \((\vec r,\vec u)\) by
\begin{equation}
        \dot{\vec r} = \grad_{u} \psi,
        \qquad
        \dot{\vec u} = -\grad_{r} \psi.
\end{equation}
Unlike in the Hamiltonian case, the corresponding phase-space flow is not, in general, incompressible.
Its divergence is
\begin{equation}
        \operatorname{div}\,(\dot{\vec \varphi}) = \{s,T\} \equiv \grad_{r} s \cdot \grad_{u} T - \grad_{u} s \cdot \grad_{r} T,
\end{equation}
so that the Poisson bracket \(\{s,T\}\) measures the local expansion or contraction of neighbouring trajectories in phase space.

The detailed geometric implications of this structure were the main focus of Paper~I.
Here, we use the same framework primarily as the starting point for the averaged equations and their one-dimensional closure.
The mean density is obtained as the velocity marginal of \(\rho\),
\begin{equation}
        \bar{\rho}(\vec{r},t)=\int \rho(t,\vec{r},\vec{u})\,d\vec{u},
\end{equation}
and the Favre average of any specific quantity \(x(t,\vec{r},\vec{u})\) is defined by
\begin{equation}
        \tilde{x}(\vec{r},t)
        =
        \frac{\displaystyle \int \rho x\,d\vec{u}}
             {\displaystyle \int \rho\,d\vec{u}}
        =
        \frac{\langle \rho x \rangle}{\bar{\rho}}.
\end{equation}

It is convenient to introduce the conditional velocity distribution
\begin{equation}
        \label{eq:f_definition}
        f(\vec u \mid \vec r, t)
        \equiv
        \frac{\rho(t, \vec r, \vec u)}{\bar\rho(\vec r, t)},
        \qquad
        \langle f \rangle = 1,
\end{equation}
which describes how the mass at a fixed position is distributed over velocities.
The Favre average of any specific quantity may then be written as an expectation under \(f\),
\begin{equation}
        \label{eq:favre_as_f}
        \tilde x = \langle f x \rangle,
\end{equation}
and the associated Favre fluctuation \(x'' \equiv x - \tilde x\) satisfies \(\langle f x'' \rangle = 0\).
In this way, a single notation \(\langle \cdot \rangle\) serves throughout for velocity-space integrals, and all Favre averages reduce to moments of \(f\).
The distribution \(f\) plays a central role in the present paper: the closure problem developed in the following sections amounts to specifying \(f\) under a limited set of mean constraints, since all transport terms entering the equations are determined by its moments.

As shown in Paper~I, the usual mean conservation laws follow directly from velocity moments of the phase-space continuity equation.
More precisely, if a specific quantity \(x\) satisfies a local balance of the form
\begin{equation}
        \label{eq:local_evolution_phase_space}
        \rho \Dt x + \grad_{r} \cdot \vec{Q}_x = \rho S_x,
        \qquad \Dt \equiv \dt + \vec u \cdot \grad_{r} + \vec \gamma \cdot \grad_{u},
\end{equation}
then the corresponding mean equation may be written as
\begin{equation}
        \label{eq:averaged_evolution_phase_space}
        \partial_t \langle \rho x \rangle
        +
        \grad \cdot
        \Big[
                \langle \rho x \rangle\,\tilde{\vec{u}}
                +
                \langle \rho x'' \vec{u}'' \rangle
                +
                \langle \vec{Q}_x \rangle
        \Big]
        =
        \langle \rho S_x \rangle,
\end{equation}
where \(\vec{Q}_x\) and \(S_x\) denote the local fluxes and sources associated with \(x\). 
In this way, the turbulent terms entering the one-dimensional description are not introduced independently, but arise as moments of the same phase-space distribution.
For instance, applying this procedure to specific enthalpy \(h = \varepsilon + pv\) in the low-Mach regime considered here gives
\begin{equation}
        \label{eq:enthalpy_conservation}
        \partial_t \langle \rho h \rangle
        + \grad \cdot
        \Big[\langle \rho h \rangle \, \tilde{\vec u}
                + \langle \rho h'' \vec u'' \rangle
                + \vec Q_r
        \Big]
        = \langle \rho S_h \rangle,
\end{equation}
where \(\vec Q_r = \langle \vec Q_h \rangle\) accounts for the radiative (and possibly conductive) transport and \(S_h\) for the local production of enthalpy by nuclear reactions and compressibility effects.

The first three moments of the continuity equation, corresponding to \(x = 1\), \(x = \vec u\) and \(x = k = \vec u^{\,2}/2\), yield the mean equations of mass, momentum, and kinetic-energy conservation,
\begin{align}
        \label{eq:mass_conservation}
        &\partial_t \bar\rho + \grad \cdot \langle \rho \vec u \rangle = 0,
        \\
        \label{eq:momentum_conservation}
        &\partial_t \langle \rho \vec u \rangle
        + \grad \cdot \Big(\langle \rho \vec u \rangle \, \tilde{\vec u} + \mathrm R\Big)
        = \langle \rho \vec\gamma \rangle,
        \\
        \label{eq:kinetic_energy_conservation}
        &\partial_t \langle \rho k \rangle
        + \grad \cdot \Big(\langle \rho k \rangle \, \tilde{\vec u} + \vec Q_k\Big)
        = \langle \rho \vec u \cdot \vec\gamma \rangle,
\end{align}
with
\begin{ceqn}
        \[
                \mathrm R = \langle \rho \vec u'' \vec u'' \rangle,
                \qquad
                \vec Q_k = \langle \rho k'' \vec u'' \rangle.
        \]
\end{ceqn}
Equations~\eqref{eq:kinetic_energy_conservation} and \eqref{eq:enthalpy_conservation} can then be combined to give the mean-field balance of total energy \(H = h + k + \Phi\), confirming that the mean-field equations emerge naturally as velocity-space moments of the phase-space continuity law.

\subsection{Stationary radial reduction}
\label{sec:stationary_radial_reduction}

The mean equations derived above hold in full generality.
For stellar envelope modelling, however, we are primarily interested in the stationary, spherically symmetric case, in which phase space reduces to the pair \((r,u)\), where \(u\) denotes the radial velocity and all mean quantities depend on \(r\) alone.
The conditional velocity distribution accordingly reduces to \(f(u \mid r)\), a one-dimensional distribution over radial velocity at a fixed radius, and the Favre average of any specific quantity \(x(r,u)\) becomes
\begin{equation}
        \tilde x(r) = \langle f x \rangle = \int f(u \mid r) \, x(r,u) \, du.
\end{equation}

For a bounded envelope with no net mass loss, the stationary form of the mass equation, Eq.~(\ref{eq:mass_conservation}), implies
\begin{equation}
        \label{eq:zero_mass_flux_bulk}
        \bar\rho \, \langle f u \rangle = 0.
\end{equation}
The Favre mean radial velocity therefore vanishes, and \(u\) may be identified with its fluctuation in the expressions that follow.
Once this condition is imposed, only two quantities retain the imprint of convective motions in the mean equations: the specific turbulent pressure,
\begin{equation}
        \label{eq:pi_def}
        \pi \equiv \langle f u^2 \rangle,
\end{equation}
which is simply the velocity variance of \(f\) and measures the kinematic spread of the local velocity population, and the specific turbulent energy flux,
\begin{equation}
        \label{eq:phi_def}
        \phi \equiv \langle f H'' u \rangle,
        \qquad H'' = h'' + k'',
\end{equation}
which represents the flux of enthalpy and kinetic energy per unit mass carried by the velocity correlations within \(f\).

These quantities enter the stationary radial equations as follows.
The momentum equation, Eq.~(\ref{eq:momentum_conservation}), reduces to
\begin{equation}
        \frac{d}{dr}\Bigl(p + \bar\rho \pi\Bigr) = -\bar\rho g,
\end{equation}
where \(g\) is the local gravitational acceleration, while the total-energy equation reads
\begin{equation}
        \label{eq:flux_balance}
        Q_r + \bar\rho \phi = Q,
\end{equation}
Here, \(Q\) is the total flux and \(Q_r = -\chi \, d\tilde T/dr\) is the radiative flux.
The closure problem is thus reduced to determining \(\pi\) and \(\phi\) from the local stratification, that is, to specifying \(f(u \mid r)\).

\subsection{Statistical entropy and formulation of the closure problem}
\label{sec:statistical_entropy}

The closure proposed below is rooted in the phase-space interpretation developed in Paper~I.
A central result of that framework is the local evolution law
\begin{equation}
        \label{eq:local_statistical_information}
    \Dt(-\ln\rho) = \{s,T\},
\end{equation}
where \(\rho\) denotes the phase-space mass density.
Since \(-\ln\rho\) may be interpreted as the statistical information carried per unit phase-space mass, the bracket \(\{s,T\}\) therefore represents both the divergence of the Liouvillian phase-space flow and the local source of statistical entropy per unit mass.
We emphasise here that this statistical information must be distinguished from the thermodynamic entropy \(s(r,u)\) of an individual fluid particle: the latter characterises its local thermodynamic state, whereas \(-\ln\rho\) characterises how the phase-space mass is distributed over the accessible velocities.

The corresponding evolution law for a population of fluid elements follows directly from this local identity.
Consider a material domain \(\mathcal{D}(t)\) transported by the phase-space flow, and denote its conserved mass and mass-specific statistical entropy by
\begin{ceqn}
        \[
            M_{\mathcal D}
            =
            \int_{\mathcal D(t)} \rho\,d\Gamma, 
            \qquad
            \Sigma_{\mathcal D}(t)
            =
            \frac{1}{M_{\mathcal D}}
            \int_{\mathcal D(t)}
            -\rho\ln\rho\,d\Gamma.
    \]
\end{ceqn}
Here \(d\Gamma\) is the phase-space volume element.
For any scalar quantity \(x\), conservation of phase-space mass \eqref{eq:phase_space_continuity} gives
\begin{equation}
    \frac{d}{dt}
    \int_{\mathcal D(t)}
    \rho x\,d\Gamma
    =
    \int_{\mathcal D(t)}
    \rho\Dt x\,d\Gamma.
\end{equation}
Applying this identity to \(x=-\ln\rho\) therefore yields
\begin{equation}
    \label{eq:statistical_entropy_material}
    \frac{d\Sigma_{\mathcal D}}{dt}
    =
    \frac{1}{M_{\mathcal D}}
    \int_{\mathcal D(t)}
    \rho\{s,T\}\,d\Gamma.
\end{equation}
Consequently, a material population evolving within a region where \(\{s,T\} > 0\) undergoes a monotonic increase of its statistical entropy.
In the convective bulk identified in Paper~I, the populated phase-space flow is predominantly expansive so that this statistical tendency applies over most of the realised support.

The same production law can be written in Eulerian form.
Multiplying Eq.~\eqref{eq:local_statistical_information} by \(\rho\) and using Eqs.~\eqref{eq:local_evolution_phase_space} and \eqref{eq:averaged_evolution_phase_space} yields
\begin{equation}
        \label{eq:statistical_entropy_eulerian}
    \partial_t(\bar\rho \Sigma)
    +
    \grad \cdot
    \bigl(
        \bar\rho \Sigma\,\tilde{\vec u}
        +
        \vec J_\Sigma
    \bigr)
    =
    \bar\rho \langle f \{s,T\} \rangle,
\end{equation}
where
\begin{ceqn}
        \[
    \Sigma
    =
    \left \langle -f\ln\rho \right \rangle,
    \qquad
    \vec J_\Sigma
    =
    \bar\rho
    \left\langle
        -f\ln\rho\,\vec u''
    \right\rangle
    \]
\end{ceqn}
denote the local statistical entropy and its non-advective transport flux, respectively.

While Eq~\eqref{eq:statistical_entropy_material} isolates the statistical-entropy production by following a closed material population, the Eulerian form \eqref{eq:statistical_entropy_eulerian} shows how this production is redistributed between neighbouring layers.
In a stationary stellar envelope with vanishing mean mass flux, \(\tilde{\vec u}=0\), and the balance reduces to
\begin{equation}
    \grad\cdot\vec J_\Sigma
    =
    \bar\rho\langle f\{s,T\}\rangle.
\end{equation}
A positive local production can therefore be sustained in a statistically stationary state through the transport of statistical entropy across the spatial domain.
This distinction becomes particularly important near the boundaries of the convective region, where exchanges between dynamically distinct populations contribute substantially to the local distribution (cf Sect.~\ref{sec:multi_population}).

For the local closure problem, the relevant functional is the statistical entropy \(\Sigma\) itself.
Using \(\rho=\bar\rho f\), it can be written at a fixed radius as
\begin{equation}
    \Sigma(r)
    =
    -\ln\bar\rho(r)
    -
    \int f(u\mid r)\ln f(u\mid r)\,du.
\end{equation}
The first term depends only on the mean density and is therefore fixed once the mean thermodynamic structure is specified.
At a given radius, maximising the statistical entropy is consequently equivalent to maximising the Shannon entropy
\begin{equation}
        \label{eq:Shannon_entropy}
    \mathcal S[f]
    \equiv
    -\int f(u\mid r)\ln f(u\mid r)\,du
\end{equation}
over velocity distributions satisfying the local bulk constraints.

The increase of statistical entropy derived above provides a physical motivation for seeking a bulk velocity distribution with high statistical entropy.
At a fixed mean stratification, this amounts to seeking the velocity distribution that maximises the Shannon entropy \(\mathcal S[f]\) while satisfying the physical constraints.
The PSA evolution law does not, however, establish that a material population necessarily evolves all the way to this maximum.
In kinetic theory, this connection is provided by the Boltzmann collision operator, whose entropy production vanishes precisely for local Maxwellian distributions.
No equivalent result has yet been established for the PSA dynamics.
The maximum-entropy distribution is therefore adopted here as a closure for the convective bulk, motivated by the direction of the statistical evolution and subsequently tested against three-dimensional simulations.

\subsection{Maximum-entropy distribution and closed moments}
\label{sec:maxent_distribution}

The maximum-entropy closure relies on two modelling assumptions motivated by the behaviour of the convective bulk discussed in Paper~I.
The first is that, over the realised part of the local velocity distribution, the thermodynamic entropy may be approximated by its first-order dependence on velocity,
\begin{equation}
        \label{eq:assumption1}
        s(r,u)\simeq \tilde s(r) + u\,\partial_u s(r).
\end{equation}
This approximation is consistent with the local quasi-adiabatic regime identified in Paper~I, in which locality holds over the acceleration time, non-advective entropy exchange remains negligible over that interval, and entropy fluctuations within the local velocity population remain moderate.
Under these conditions, the velocity--entropy relation retains a positive slope over the realised support, with \(\partial_u s>0\) to leading order, consistently with the predominantly expansive phase-space flow \(\{s,T\}>0\) identified in the convective bulk of Paper~I.

The second assumption is the existence of a deep entropy level \(S\), set by the underlying stellar interior and inherited by rising fluid elements in the nearly adiabatic convective bulk.
In an unstable region, these rising elements encounter, on average, material with an entropy deficit relative to \(S\), so that radiative exchange induces a net heat loss along their motion and \(\Dt s\leq 0\).
Their entropy is therefore reduced below, or at most maintained at, the value inherited from below.
The descending elements are then surrounded by fluid with \(s\leq S\); radiative diffusion may bring them back towards this value, but cannot raise them above it.
Thus the realised convective states satisfy
\begin{ceqn}
        \[
                s(r,u)\le S.
        \]
\end{ceqn}
Combined with the linear approximation above, this condition implies an upper bound in velocity space,
\begin{ceqn}
        \[
                u\le U(r) \equiv \frac{S-\tilde s(r)}{\partial_u s(r)}.
        \]
\end{ceqn}

At a fixed radius, the closure therefore amounts to finding the velocity distribution of maximum statistical entropy on the half-line \(]-\infty,U]\), under the constraint of vanishing mean mass flux.
Introducing the shifted distribution
\begin{equation}
        g(w \mid r)\equiv f(U-w \mid r),
        \qquad
        w = U-u \in \mathbb{R}^+,
\end{equation}
maps the problem onto the positive half-line.
The vanishing mean mass flux condition, \(\langle f u \rangle = 0\), then becomes
\begin{equation}
        \label{eq:constraint_mean_g}
        \langle g w \rangle = U.
\end{equation}
The problem is therefore reduced to finding the distribution on \(\mathbb{R}^+\) with a prescribed mean \(U(r)\) that maximises \(\mathcal{S}[g]\) (see Eq.~\ref{eq:Shannon_entropy}).
Introducing the Lagrange multipliers \(\lambda\) for normalisation and \(\mu\) for the mean, one requires the stationarity of
\begin{equation}
        \mathcal{L}[g]
        =
        \mathcal{S}[g]
        +\lambda \, \bigl (1 - \langle g \rangle \bigr )
        +\mu \, \bigl (U - \langle g w \rangle \bigr).
\end{equation}
For arbitrary variations \(\delta g\), this gives
\begin{equation}
        \delta \mathcal{L}
        =
        - \,
        \Bigl \langle
                \bigl(\ln g + 1 + \lambda + \mu w\bigr)\,\delta g
        \Bigr \rangle
        =0,
\end{equation}
which requires the factor \(\ln g + 1 + \lambda + \mu w\) to vanish pointwise and therefore
\begin{ceqn}
        \[
                g(w \mid r)\propto e^{-\mu(r) w}.
        \]
\end{ceqn}
Normalisation together with the prescribed mean gives
\begin{equation}
        g_*(w \mid r)
        =
        \frac{1}{U(r)}
        \exp\left[-\frac{w}{U(r)}\right].
\end{equation}

This solution is also the unique maximum of the statistical entropy under the imposed constraints.
Indeed, since
\begin{equation}
        \ln g_*
        =
        -\ln U-\frac{w}{U},
\end{equation}
normalisation and the common mean \(\langle w\rangle=U\) imply
\begin{equation}
        \langle g\ln g_*\rangle
        =
        \langle g_*\ln g_*\rangle.
\end{equation}
The Kullback--Leibler divergence from \(g_*\) \citep{Kullback1951} can therefore be written as
\begin{equation}
        D_{\rm KL}(g\Vert g_*)
        =
        \bigl\langle
        g\ln (g/g_*)
        \bigr\rangle
        =
        \mathcal{S}[g_*]-\mathcal{S}[g]
        \geq0.
\end{equation}
Hence \(\mathcal{S}[g]\leq\mathcal{S}[g_*]\), with equality only for \(g=g_*\).
Returning to the original velocity variable gives the unique maximum-entropy distribution
\begin{equation}
        \label{eq:exp_distribution_bulk}
        f(u \mid r)
        =
        \frac{1}{U(r)}
        \exp \left(\frac{u}{U(r)}-1\right),
        \qquad
        u\le U(r).
\end{equation}
The resulting distribution is asymmetric, reflecting the one-sided support inherited from the entropy bound.

Once Eq.~\eqref{eq:exp_distribution_bulk} is known, the moments entering the one-dimensional mean equations follow immediately.
Since \(w = U-u\) is exponentially distributed with mean \(U\), its raw moments satisfy
\begin{equation}
        \langle g w^n \rangle = n!\,U^n.
\end{equation}
Using \(u=U-w\), one finds in particular
\begin{equation}
        \label{eq:f_moments}
        \langle f u \rangle = 0,
        \qquad
        \pi = \langle f u^2 \rangle = U^2,
        \qquad
        \langle f u^3 \rangle = -2U^3,
\end{equation}
where \(\pi\) is the specific turbulent pressure introduced in Eq.~\eqref{eq:pi_def}.

For the specific turbulent energy flux~\eqref{eq:phi_def}, the enthalpy fluctuation must be related to the velocity.
Under the present low-Mach assumptions, pressure fluctuations are neglected to leading order, so that \(h''\simeq\tilde T s''\).
Together with the first assumption~\eqref{eq:assumption1}, this gives
\begin{equation}
        \label{eq:local_enthalpy_flux}
        h''u
        \simeq
        \tilde T s''u
        \simeq
        \tilde T u^2\partial_u s.
\end{equation}
Introducing the local entropy contrast
\begin{equation}
        \delta s \equiv S-\tilde s = U\,\partial_u s,
\end{equation}
the mean enthalpy flux becomes
\begin{equation}
        \langle f h''u\rangle
        \simeq
        \tilde T\partial_u s\,\langle f u^2\rangle
        =
        \tilde T\delta s\,U.
\end{equation}
Combining this result with the third moment of \(f\) obtained in Eq.~\eqref{eq:f_moments}, Eq.~\eqref{eq:phi_def} yields
\begin{equation}
        \label{eq:phi_exponential}
        \phi = \tilde T\delta s U-U^3.
\end{equation}
The first term is the enthalpy contribution, proportional to the entropy contrast \(\delta s\) and to the upper velocity bound \(U\).
The second is an oppositely directed kinetic-energy contribution proportional to \(U^3\).
For a given value of \(\phi\), Eq.~\eqref{eq:phi_exponential} still admits a continuous family of pairs \((\tilde T\delta s,U)\).
The next step is therefore to determine how the convective envelope selects the relation between the entropy contrast \(\delta s\) and the velocity scale \(U\) for a prescribed energy flux.

\subsection{Convective optimum and thermal stratification}
\label{sec:convective_optimum}

Such a principle is needed because states compatible with a given energy flux \(\phi\) are not equivalent from the point of view of the mean structure.
A larger entropy contrast \(\delta s\) corresponds to a larger departure from the deep adiabat and therefore to a more substantial structural adjustment, whereas variations in \(U\) primarily reflect changes in the kinematic spread of the velocity distribution.
In that sense, \(\tilde T \delta s\) may be viewed as the structural price paid by the envelope in order to sustain a given turbulent transport.
It then becomes natural to select the state for which this price is minimal.
At a fixed local thermodynamic scale \(\tilde T \delta s\), the same idea may be reformulated by asking which maximum upper velocity \(U\) allows the distribution to carry the largest convective flux.
We adopted this selection principle here and refer to the resulting state as the convective optimum.

The maximum of \(\phi(\tilde T \delta s, U)\) at a fixed \(\tilde T \delta s\) is obtained from
\begin{equation}
        \left(\frac{\partial \phi}{\partial U}\right)_{\tilde T\delta s}
        =
        \tilde T \delta s - 3U^2
        = 0,
\end{equation}
so that the optimum satisfies
\begin{equation}
        \label{eq:optimum_relation_section2}
        \tilde T \delta s = 3U^2.
\end{equation}
Substituting this relation into Eq.~\eqref{eq:phi_exponential} gives
\begin{equation}
        \label{eq:phi_optimum}
        \phi = 2U^3.
\end{equation}
The selected maximum upper velocity and entropy contrast may therefore be written directly as functions of \(\phi\),
\begin{equation}
        \label{eq:optimum_section2}
        U = \left(\frac{\phi}{2}\right)^{1/3},
        \qquad
        \tilde T \delta s
        =
        3\left(\frac{\phi}{2}\right)^{2/3}.
\end{equation}

This result admits a simple physical interpretation.
At a fixed entropy contrast, the enthalpy flux in Eq.~\eqref{eq:phi_exponential} grows linearly with \(U\), whereas the oppositely directed kinetic-energy flux grows as \(U^3\).
The optimum therefore corresponds to the point beyond which increasing the maximum upper velocity no longer improves the net transport efficiently, because the kinetic contribution cancels an increasing fraction of the enthalpy flux.
In that sense, the convective optimum selects the most efficient state compatible with the maximum-entropy closure derived above.

An important consequence of this optimum is that, once \(\phi\) is known, the full thermal structure follows without any additional free parameter.
Since \(S\) is constant in the present closure, the definition \(\delta s = S - \tilde s\) implies \(d\tilde s = -d(\delta s)\).
Introducing the usual temperature gradients \(\nabla\) and \(\DTad\) as well as the dimensionless entropy contrast
\begin{equation}
        \zeta
        \equiv
        \frac{\delta s}{c_p}
        =
        \frac{3}{c_p\tilde T}
        \left(\frac{\phi}{2}\right)^{2/3},
\end{equation}
the thermodynamic relation \(d\tilde s = c_p(d\ln\tilde T - \DTad \, d\ln p)\) gives
\begin{equation}
        \label{eq:nabla_intermediate_section2_compact}
        \nabla
        =
        \DTad
        -
        \zeta \, \frac{d\ln\delta s}{d\ln p}.
\end{equation}
Using Eq.~\eqref{eq:optimum_section2}, namely \(\delta s \propto \phi^{2/3} / \tilde T\), one may then isolate \(\nabla\) to obtain
\begin{equation}
        \label{eq:nabla_optimum_section2}
        \nabla
        =
        (1-\zeta)^{-1}
        \left(
        \DTad
        -
        \frac{2\zeta}{3}\frac{d\ln\phi}{d\ln p}
        \right).
\end{equation}
In the present single-population framework, the flux balance~\eqref{eq:flux_balance} identifies \(\phi=(Q-Q_r)/\bar\rho\), so that the thermal structure is fully determined.
In the deep efficient-convection limit, \(\delta s\to0\), and hence \(\zeta\to0\), recovering \(\nabla\to\DTad\) as expected.

This expression also raises a natural question, namely what happens as \(\zeta\) approaches unity.
To interpret this limit, it is useful to return to the definition of \(\zeta\) within the convective optimum,
\begin{equation}
        \zeta
        =
        \frac{3U^2}{c_p\tilde T}
        =
        \frac{3(\Gamma_3 - 1)}{\nu_p} \, \mathcal M^2,
        \qquad
        \mathcal M \equiv \frac{U}{c_s},
\end{equation}
with \(\nu_p \equiv (\partial\ln v / \partial\ln T)_p\).
Thus \(\zeta\) is controlled by the squared convective Mach number, modulated by the local thermodynamic response.
While in the ideal-gas limit \(\zeta=2\mathcal M^2\), it is worth noting that partial ionisation can substantially reduce the prefactor in stellar surface layers.
Seen in this light, the condition \(\zeta < 1\) is more than a mere mathematical restriction of the formula.
It indicates the range over which the assumptions underlying the model remain physically meaningful.
A convective flow approaching Mach numbers of order unity would already be entering a regime in which pressure fluctuations acquire a more dynamical role, acoustic coupling can no longer be neglected, and compressible effects such as buoyancy braking are expected to modify the transport itself \citep{Toomre1976,Latour1981}.
The formal divergence of Eq.~\eqref{eq:nabla_optimum_section2} as the flow approaches the edge of the low-Mach regime should therefore be read as the sign that the approximations from which it was derived have reached their limit of applicability.

\section{Phenomenology of 3D envelope convection and empirical constraints}
\label{sec:phenomenology_3D}

\subsection{Three-dimensional reference models}
\label{sec:3D_models}

\begin{figure}
        \centering
        \includegraphics[width=\columnwidth]{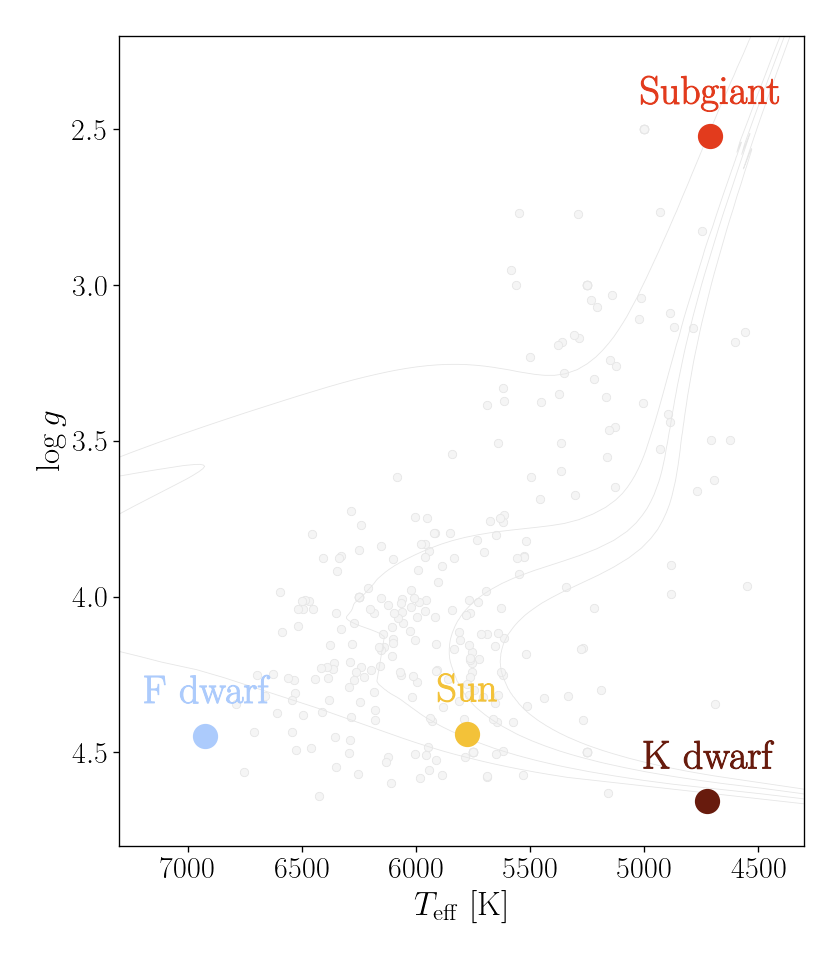}
        \caption{Location of the four reference simulations in the \((T_{\rm eff},\log g)\) plane.
        The grey points show the models available in the public M3DIS grid, while the four solar-metallicity simulations analysed in this work are highlighted.
        The thin grey curves show solar-metallicity Dartmouth isochrones at 1, 4, 8, and 12~Gyr, from left to right \citep{Dotter2008}.
        The models selected for the present study are indicated in colour with their names.}
        \label{fig:m3dis_grid}
\end{figure}

To identify which aspects of the envelope structure are already captured by the one-dimensional maximum-entropy closure derived in Sect.~\ref{sec:1D_closure}, and which require additional thermodynamic structure, we compare its predictions with three-dimensional radiation-hydrodynamical models drawn from the publicly available M3DIS grid \citep{Eitner2024}.
These local box simulations solve the compressible hydrodynamical equations together with non-grey radiative transfer, including a tabulated equation of state and opacity binning following \citet{Nordlund1982}, with opacity data based on the MARCS framework \citep{Gustafsson2008,Eitner2024}.
They therefore retain the thermodynamic and radiative processes required to study the feedback of surface convection on the mean stratification.

All four models considered here have solar metallicity.
Their location within the \((T_{\rm eff},\log g)\) coverage of the public M3DIS grid is shown in Fig.~\ref{fig:m3dis_grid}.
The sample was deliberately kept small enough for a detailed thermodynamic analysis while probing as wide a range of envelope conditions as possible.
In addition to the solar reference case, three atmospheres were selected at distinct extrema of the solar-metallicity grid: a cool high-gravity dwarf, hereafter K dwarf, \((\log g=4.66,\ T_{\rm eff}=4725\,\mathrm{K})\), which samples the highest surface gravity; a hot dwarf, hereafter F dwarf, \((\log g=4.45,\ T_{\rm eff}=6923\,\mathrm{K})\), which samples the highest effective temperature; and a low-gravity subgiant \((\log g=2.52,\ T_{\rm eff}=4712\,\mathrm{K})\), which samples the lowest surface gravity.
Together with the Sun \((\log g=4.44,\ T_{\rm eff}=5777\,\mathrm{K})\), these models provide contrasting test cases for the dependence of the mean stratification and surface transition on stellar parameters.

The simulations use a geometric vertical coordinate \(z\), with typical resolutions of order \((n_z,n_y,n_x)\sim(80\text{--}100,160\text{--}200,160\text{--}200)\).
For each model, the statistics below are constructed from 10 to 20 snapshots taken after the initial relaxation phase and averaged horizontally and temporally.
In practice, Reynolds averages are approximated by horizontal-time averages at a fixed layer, while Favre averages are constructed in the corresponding density-weighted form.
When two-dimensional distributions are shown, the colour scale represents the corresponding sampling of \(f\) in the simulation: the mass falling in a bin is normalised by the mass of the parent horizontal layer, consistently with the local interpretation \(f=\rho/\bar\rho\).
Depending on the diagnostic considered, we used either the native geometric coordinate \(z\) or derived vertical coordinates such as \(\log p\).
The former is useful for selecting a representative geometrical layer, while the latter provides a more transparent view of the thermodynamic stratification and of its connection with one-dimensional envelope profiles.
Entropies are expressed in units of \(\mathcal R=k_{\rm B}/m_{\rm u}\) and shifted by the deep adiabatic level \(S\), so that \(S\) appears at \(s=0\).

\subsection{Solar bulk structure and surface transition}
\label{sec:solar_bulk}

\begin{figure}[!ht]
    \centering
    \includegraphics[width=\columnwidth]{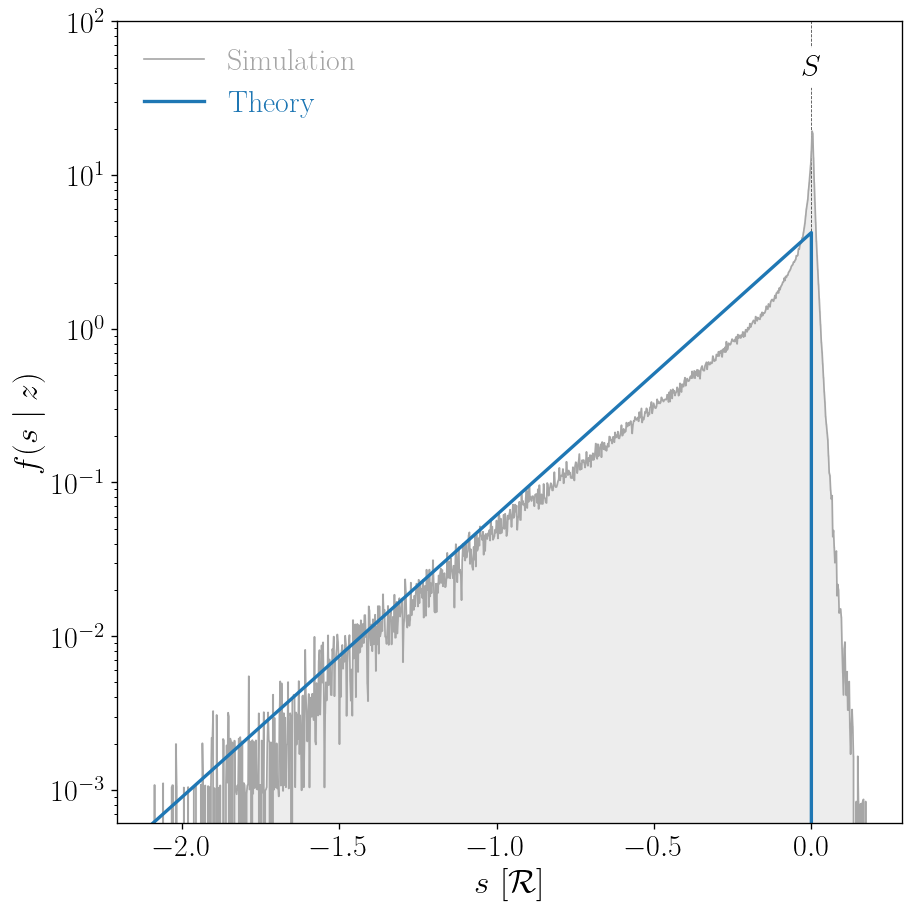}
        \caption{
        Mass-weighted entropy distribution in a representative layer of the solar convective region, located at \(z\simeq -1\,{\rm Mm}\).
        The entropy is measured relative to the deep adiabatic level \(S\), marked by the vertical dashed line.
        The grey curve shows the simulated distribution, normalised as a density in entropy.
        The blue curve shows the maximum-entropy prediction derived in Sect.~\ref{sec:1D_closure}, evaluated with the local value of \(\delta s\).
        }
    \label{fig:bulk_entropy_distribution}
\end{figure}

\begin{figure*}[!ht]
    \centering
    \includegraphics[width=\textwidth]{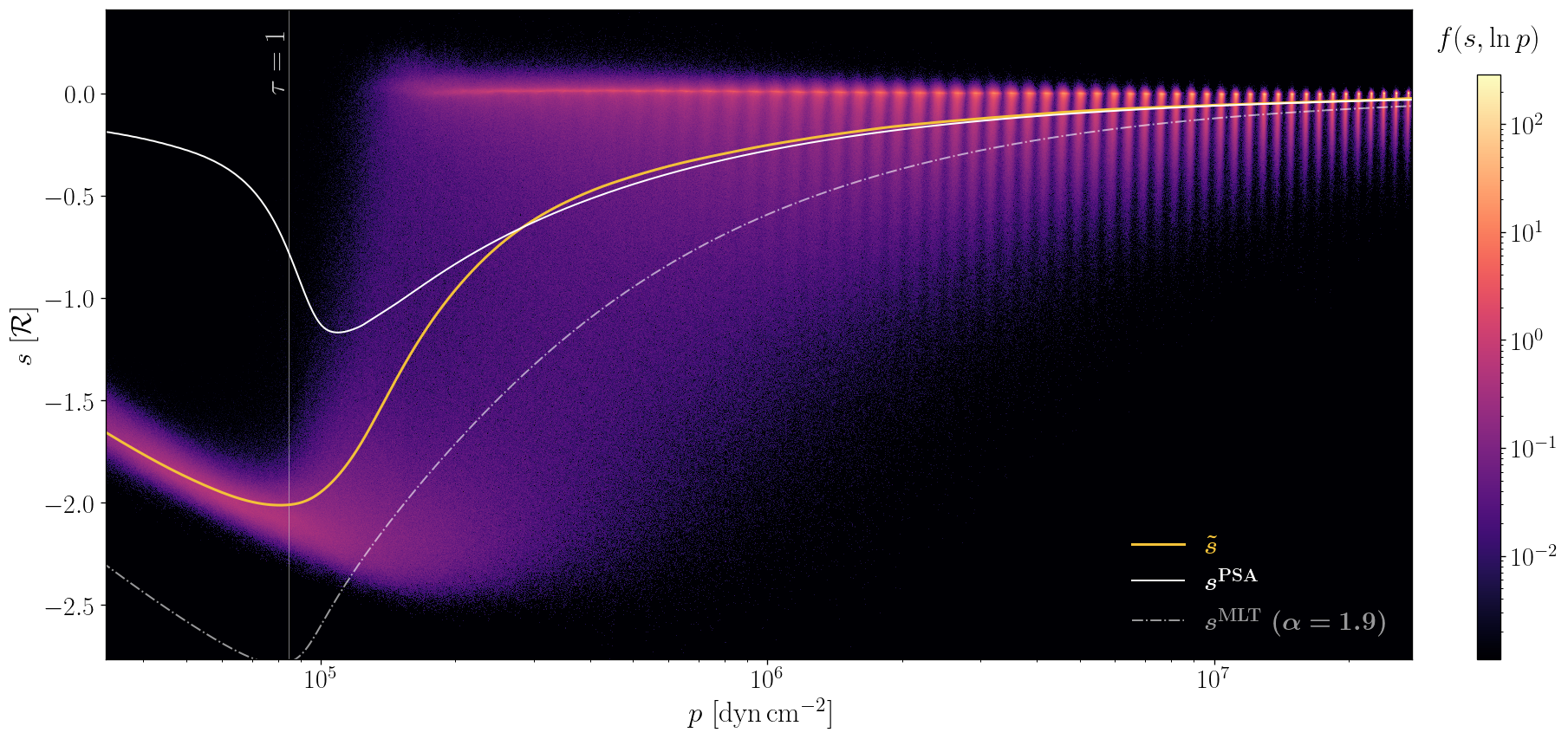}
        \caption{
        Entropy distribution of the solar simulation in the \((\ln p,s)\) plane.
        The colour scale represents the sampled distribution \(f(s,\ln p)\) in the simulation, with the entropy measured relative to the deep adiabatic level \(S\).
        The solid orange line represents the Favre-mean entropy profile \(\tilde{s}\), the light solid line shows the maximum-entropy prediction \(S-\delta s\), and the dash-dotted line indicates the calibrated mixing-length profile \(s^{\mathrm{MLT}}\) with \(\alpha=1.9\).
        The vertical line marks the approximate location of the photosphere, \(\tau_{\rm R}=1\).
        }
    \label{fig:hist_entropy}
\end{figure*}

We begin with the solar model, which provides a convenient reference for interpreting the behaviour found across the other atmospheres.
Before considering mean profiles, we examine the underlying thermodynamic distributions, which retain information that averaging necessarily compresses: they show the local shape of the entropy distribution, the presence of sharp bounds or extended tails, and eventually the coexistence of distinct thermodynamic components.
This provides a direct test of where the maximum-entropy description applies and where the surface transition introduces a different statistical organisation.

\subsubsection{Statistical evolution of the entropy distribution}

We first isolated the statistical structure of the convective bulk.
Figure~\ref{fig:bulk_entropy_distribution} shows a representative solar layer at \(z\simeq-1\,{\rm Mm}\), sufficiently below the immediate radiative transition for the bulk organisation to remain dominant.
The simulated distribution has the one-sided structure anticipated by the maximum-entropy construction of Sect.~\ref{sec:1D_closure}: it is sharply bounded near the deep adiabatic entropy \(S\) and extends towards lower entropy through an approximately exponential tail.
The theoretical distribution, evaluated with the local value of \(\delta s\), closely reproduces the slope of this tail.
The main difference is an excess concentration near the upper edge, whose possible origin is discussed in Sect.~\ref{sec:discussion}.

Figure~\ref{fig:hist_entropy} follows this distribution through the solar stratification in the \((\ln p,s)\) plane.
The pressure coordinate resolves the surface transition, while the colour scale gives the layer-normalised sampling of \(f(s,\ln p)\).
The faint vertical striations at large pressure arise from the finite vertical grid, where the pressure spread within a horizontal layer becomes small compared with the spacing between adjacent layers.

Over most of the deep convective region, the high-entropy edge remains sharply defined and the realised states stay strongly concentrated near it.
This behaviour is naturally understood if the dominant ascending motions rise nearly adiabatically from the deep envelope, carrying an entropy level upwards that remains almost unchanged until the surface cooling region is reached.
The lower-entropy tail then represents the spread of entropy deficits within this one-sided bulk distribution.
The structure changes sharply beneath the surface, where a second mode develops at lower entropy and separates from the bulk branch over less than a pressure scale height.
This reorganisation coincides with the region of strong radiative cooling and signals the emergence of a thermodynamic component not represented by the single bulk distribution.

\subsubsection{Comparison with one-dimensional entropy profiles}

Three profiles are superposed in Fig.~\ref{fig:hist_entropy}, and each must be read in a slightly different way.
The Favre-averaged entropy profile \(\tilde s\) is obtained from the simulation.
In the deep interior, where the distribution remains strongly concentrated near its upper edge, this mean profile stays close to the dominant high-entropy branch.
Near the surface, however, its interpretation becomes less straightforward.
Once the distribution becomes bimodal, the mean entropy no longer follows a single populated branch of the histogram, but lies instead between the high-entropy branch and the newly formed low-entropy branch.
While its minimum mainly reflects the strength of the radiative entropy loss, the subsequent shape of the profile seems mostly controlled by the changing relative weights of the two branches.

The maximum-entropy prediction \(S-\delta s\) is given by Eq.~\eqref{eq:optimum_section2} together with the flux balance of Eq.~\eqref{eq:flux_balance}.
In the present comparison, the radiative flux \(Q_r\) is evaluated from the corresponding one-dimensional radiative-transfer solution, since the diffusive approximation is no longer adequate in the outer layers.
With the entropy reference used here, \(S=0\), and the predicted profile therefore reduces to \(-\delta s\).

For comparison, we also evaluate the superadiabatic excess predicted by MLT along the mean thermodynamic stratification of the simulation.
At each depth, the standard cubic MLT relation in the formulation of \citet{Maeder2009} is evaluated using the simulated mean state and the same imposed total flux \(Q\).
The resulting superadiabatic excess,
\begin{equation}
        \Delta\nabla_\alpha
        \equiv
        \nabla(\bar\rho,\tilde T;\alpha)-\nabla_{\rm ad},
\end{equation}
is then integrated according to
\begin{equation}
        s^{\rm MLT}
        =
        S-\int c_p\,\Delta\nabla_\alpha\,d\ln p.
\end{equation}
The integration constant is chosen so that the profile is anchored to the same deep-entropy level \(S\) as the simulation and the maximum-entropy prediction.
The mixing-length parameter is fixed to \(\alpha=1.9\), adopted as a representative solar-calibrated value broadly consistent with dedicated calibrations for solar or near-solar conditions, although the precise value depends on the adopted MLT formulation, boundary conditions, and input microphysics \citep{Ludwig1999,Trampedach2014,Sonoi2019}.
No adjustment of \(\alpha\) is made to reproduce either the entropy minimum or the total entropy jump measured in the simulation, which therefore remain predictions of this MLT calibration.

For this solar-calibrated value of \(\alpha\), the resulting profile exhibits both a deep high-entropy region and a pronounced entropy decrease near the surface.
Its stratification nevertheless remains systematically offset from that of the simulation throughout the sub-surface layers.
The entropy drop develops more rapidly than in the simulation while the histogram is still dominated by its upper branch, and more slowly once the low-entropy component begins to emerge.
The discrepancy is therefore not limited to the total amplitude of the entropy jump, but also affects how this entropy contrast is distributed with depth.

In contrast, despite the simplicity of the expression \(S-\delta s\), the maximum-entropy prediction closely follows the simulated mean entropy throughout most of the sub-surface region.
Its departure from the simulation above the surface is not surprising: once the atmosphere is reached, the assumptions underlying the maximum-entropy closure no longer apply, and the exponential distribution ceases to provide a meaningful description of the local thermodynamic state.
More important is the fact that the divergence already begins beneath the surface, before the atmospheric regime is fully established.
This change of behaviour coincides with the appearance of the second low-entropy mode visible in Fig.~\ref{fig:hist_entropy} and points towards the emergence of a thermodynamic component outside the single exponential distribution.

This distinction is not merely of local interest.
In one-dimensional stellar models, the surface layers usually fix the outer boundary of the thermal stratification, and even a modest entropy mismatch in that region may propagate inwards and shift the deep adiabatic level itself.
Capturing the onset of this second component is therefore essential for the surface transition and for the deep adiabat selected by the envelope.

\subsubsection{Velocity distribution and characteristic bulk speed}

\begin{figure*}[!t]
    \centering
    \includegraphics[width=\textwidth]{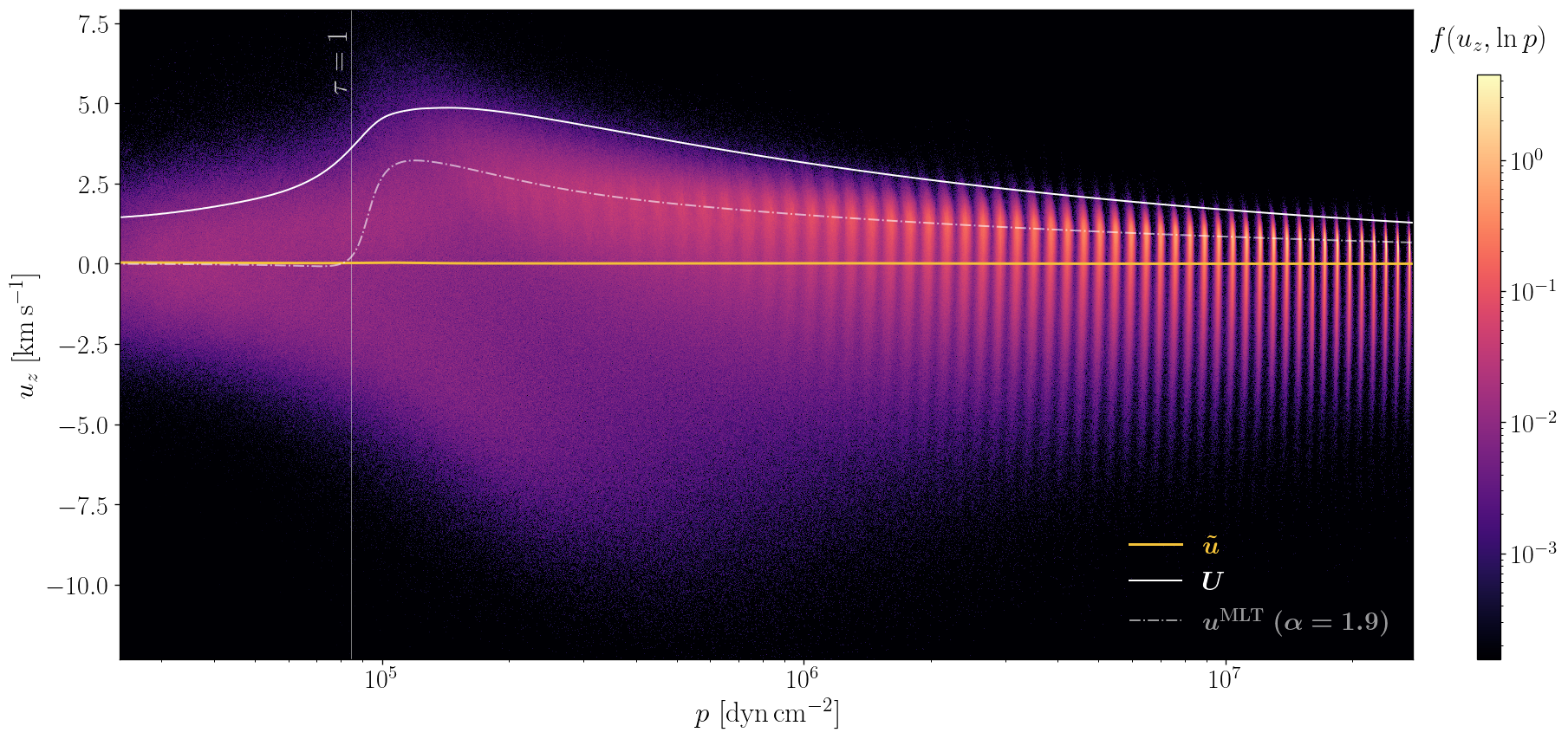}
        \caption{
        Vertical-velocity distribution of the solar simulation in the \((\ln p,u_z)\) plane.
        The colour scale represents the sampled distribution \(f(u_z,\ln p)\) in the simulation.
        The solid~yellow line shows the Favre-averaged vertical velocity \(\tilde{u}_z\), which remains close to zero; the light solid line indicates the maximum upper velocity \(U\) predicted by the maximum-entropy closure; and the dash-dotted line represents the MLT velocity \(u^{\mathrm{MLT}}\) obtained with \(\alpha=1.9\).
        The vertical line marks the approximate location of the photosphere, \(\tau_{\rm R}=1\).
        }
    \label{fig:hist_velocity}
\end{figure*}

Figure~\ref{fig:hist_velocity} shows the corresponding vertical-velocity distribution in the \((\ln p,u_z)\) plane.
The Favre-averaged vertical velocity vanishes throughout, as required by mass conservation.
The distribution itself is nevertheless strongly asymmetric: the upflows occupy a broader fraction of the volume but remain comparatively slow, whereas the downflows are less spatially extended and reach substantially larger speeds.

The upper velocity bound \(U\) predicted by the maximum-entropy closure naturally corresponds to the upper edge of the positive-velocity distribution.
Because the simulation does not satisfy the assumed entropy--velocity relation as an exact deterministic constraint, this velocity edge is less sharply defined than its entropy counterpart.
Nevertheless, throughout the bulk-dominated region, the predicted \(U\) remains close to the fastest populated ascending motions.

The MLT velocity is obtained with the same formulation and \(\alpha=1.9\) used for the entropy comparison above.
Because the MLT formalism does not describe a velocity distribution, but only a single characteristic speed, the physical interpretation of this quantity is not fixed a priori.
In practice, the comparison suggests that \(u^{\mathrm{MLT}}\) corresponds neither to the upper envelope of the histogram nor to any mean vertical velocity, which vanishes by construction, but rather to a typical convective speed close to the mode of the distribution.
In that sense, the MLT velocity retains the correct order of magnitude, but it does not encode the same information as \(U\), which is instead tied to the upper edge of the dominant ascending branch.

Finally, the velocity histogram also shows the kinematic counterpart of the second mode identified in entropy.
In the upper convective layers, an additional component becomes visible within the negative-velocity side of the distribution, suggesting that the process responsible for this structure acts asymmetrically on rising and sinking motions.

\subsection{Radiative organisation of the surface transition}

The solar histograms discussed above show that the maximum-entropy closure remains accurate while the distribution is organised around a single dominant branch.
The first clear departure from this regime coincides with the appearance of a low-entropy contribution beneath the surface.
This suggests that the relevant question is how the radiative transition reorganises the flow into distinct thermodynamic components.

\subsubsection{Radiative constraint on the ascending branch}

\begin{figure*}[!t]
    \centering
    \includegraphics[width=\textwidth]{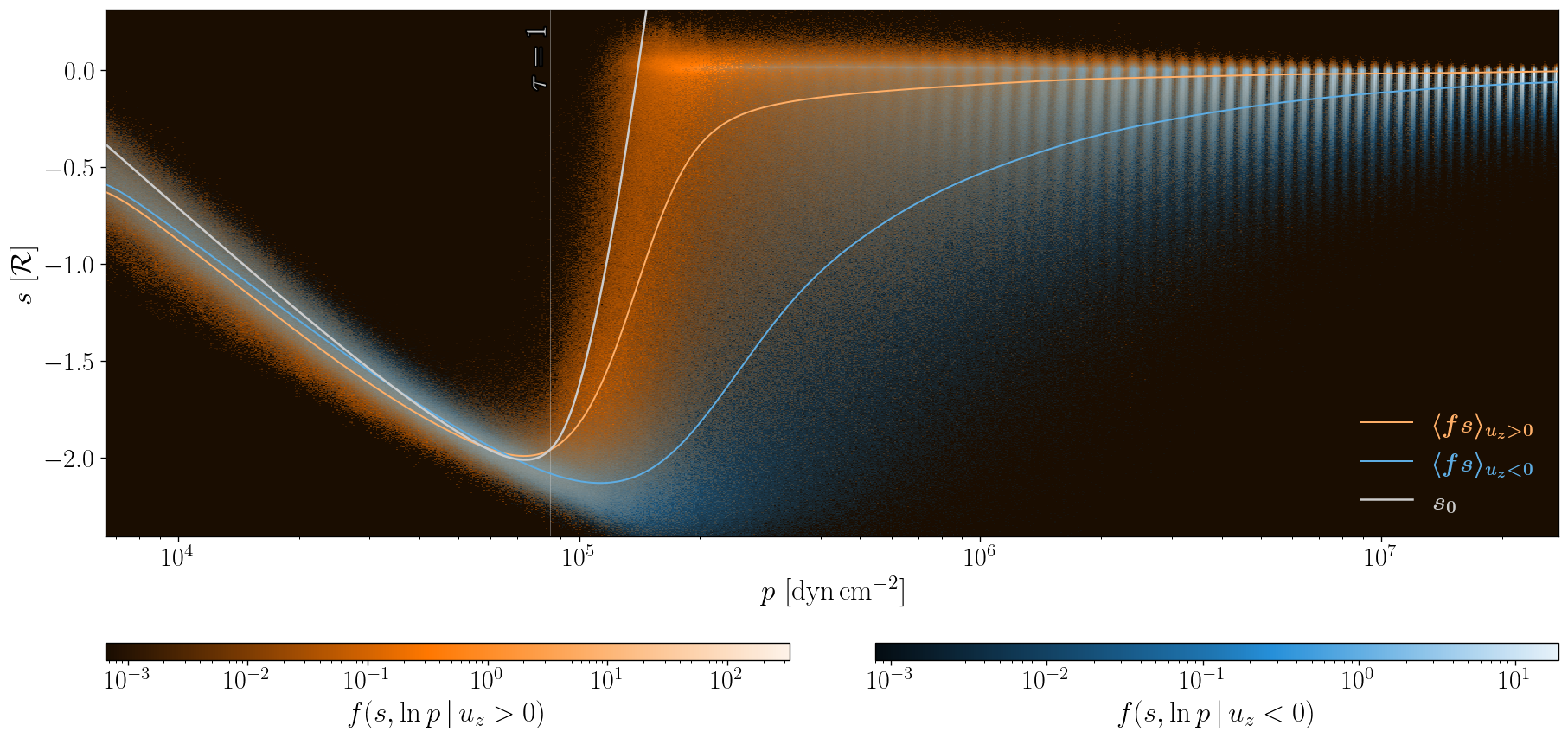}
        \caption{
        Entropy distribution of the solar simulation in the \((\ln p,s)\) plane, separated according to the sign of the vertical velocity.
        Upflows (\(u_z>0\)) are shown in orange and downflows (\(u_z<0\)) are in blue.
        The solid coloured lines show the corresponding conditional means, \(\langle f s \rangle_{u_z>0}\) and \(\langle f s \rangle_{u_z<0}\), computed with the same layer-normalised mass sampling as in Fig.~\ref{fig:hist_entropy}.
        The white curve shows the radiative entropy branch \(s_0\) constructed from the grey Eddington reference profile, and the vertical line marks \(\tau_{\rm R}=1\).
        }
    \label{fig:hist_entropy_cond}
\end{figure*}

Figure~\ref{fig:hist_entropy_cond} shows the entropy distribution of the solar simulation in the \((\ln p,s)\) plane, with upflows and downflows represented separately.
In the deep convective region, the ascending flow remains concentrated near a narrow high-entropy branch, while the descending flow occupies a broader distribution extending towards lower entropy.
This asymmetry is consistent with the one-sided bulk distribution discussed in Sect.~\ref{sec:1D_closure}.
The qualitative change occurs closer to the surface, where the ascending branch bends sharply downwards, and once the atmosphere is reached, it remains confined around a well-defined branch.

To identify the nature of that branch, it is useful to compare the distribution with a simple radiative reference.
We used the grey Eddington temperature profile
\begin{equation}
        {T_0}^4(\tau)
        =
        \frac{3}{4}T_{\rm eff}^4
        \left(\tau+\frac{2}{3}\right),
\end{equation}
where \(\tau\) denotes the Rosseland optical depth.
This profile represents the temperature branch of a grey layer carrying the total stellar flux \(Q\) radiatively, and we verified that replacing it by a numerical one-dimensional radiative-transfer profile produces no appreciable change in the resulting entropy branch for the simulations considered here.
Along the mean pressure stratification of the simulation, it defines the corresponding entropy branch \(s_0\) through
\begin{equation}
        \label{eq:entropy_radiative_branch}
        ds_0
        =
        c_p (d \ln T_0 - \DTad \, d \ln p),
\end{equation}
with the additive constant fixed by the entropy reference used in the figure.

The comparison gives a clear interpretation of the warm atmospheric branch.
The upper edge of the ascending flow is rapidly brought into contact with \(s_0\) and then remains closely attached to it.
In that sense, \(s_0\) acts as an upper radiative ceiling for the ascending flow.
A fluid element whose entropy exceeds this branch is out of balance with the grey radiative state carrying the imposed stellar flux and is therefore driven back towards it by net radiative cooling.
The same statement may be understood globally: maintaining the warm branch above \(s_0\) would require a radiative flux larger than the imposed flux, unless compensated by an oppositely directed convective contribution.
Thus \(s_0\) provides the highest radiatively self-consistent entropy branch available to the ascending material, explaining why the warm component remains closely pinned to it once it reaches the atmosphere.

\subsubsection{Generation and evolution of the plume branch}

Conditioning the entropy distribution on the sign of \(u_z\) also identifies the low-entropy component suggested by Fig.~\ref{fig:hist_entropy}.
This component is carried predominantly by the descending flow.
Its evolution differs from that of the ascending branch: after the surface transition has been crossed, the downflow branch departs from \(s_0\), develops a broad entropy deficit just beneath the photosphere, and then relaxes only gradually towards the deep bulk distribution over a much larger pressure range.
The contrast between the two branches is therefore one of thermodynamic history as much as one of mean entropy level.
The ascending flow is rapidly adjusted to the radiative branch, whereas the descending flow retains the imprint of a past cooling event far below the surface \citep{Rieutord1995}.

This behaviour is naturally interpreted as the signature of plume production by surface radiative cooling.
In realistic simulations of solar and stellar surface convection, the flow is driven primarily by entropy losses in a thin radiative boundary layer, which generate negatively buoyant material and feed the descending plumes \citep{Stein1998,Nordlund2009}.
In entropy space, this history appears as a distinct branch: it is created near the radiative transition, reaches its largest entropy contrast slightly below it, and is then progressively mixed into the deep convective bulk.

The surface transition thus appears as a structured reorganisation into a radiatively constrained ascending branch and a distinct cooled descending population.
This coupled upflow--plume structure replaces the single-branch bulk description in the near-surface layers.

\subsection{Statistical transition towards a multi-population description}
\label{sec:transition_multipop}

\begin{figure}[!t]
    \centering
    \includegraphics[width=\columnwidth]{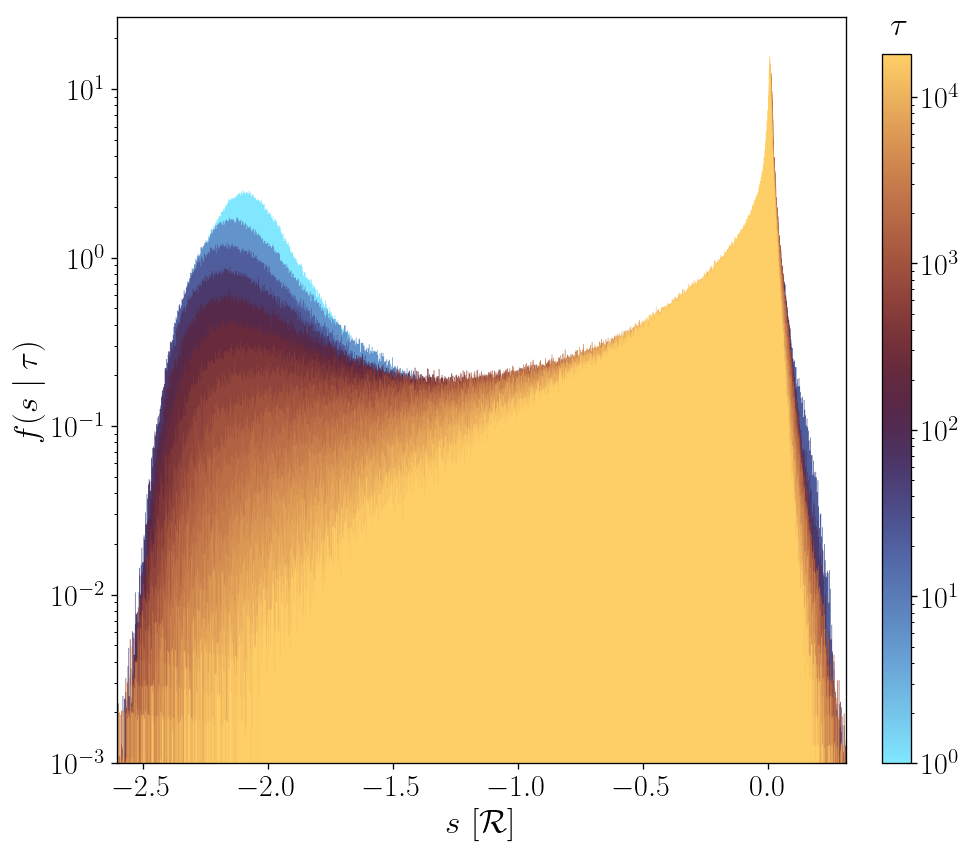}
    \caption{
    Mass-weighted entropy distributions measured at several optical depths in the solar simulation.
    Each curve is normalised as a density in entropy and is coloured by the corresponding Rosseland optical depth \(\tau\).
    }
    \label{fig:entropy_slices}
\end{figure}

The conditioned histograms shown above identify the radiative origin of the two branches.
A complementary view is obtained by returning to the local entropy distributions introduced in Fig.~\ref{fig:bulk_entropy_distribution} and following how their shape changes with optical depth.
Figure~\ref{fig:entropy_slices} shows mass-weighted entropy distributions for layers spanning \(1 \leq \tau \leq 20\,000\), each normalised to unit integral.

The deepest curve, at \(\tau \simeq 20\,000\), corresponds approximately to the bulk layer isolated in Fig.~\ref{fig:bulk_entropy_distribution}.
As the surface is approached, the high-entropy side remains strongly constrained while the lower-entropy part of the distribution changes progressively.
An additional contribution develops below the main branch.
Compared with the extended one-sided tail of the deep bulk distribution, this feature is narrower and more nearly symmetric about its own maximum, and becomes increasingly distinct as it grows.

This change of shape is the point that matters for the closure problem.
Figure~\ref{fig:entropy_slices} suggests a more structured reorganisation than a continuous deformation of the bulk distribution: a low-entropy component appears near the radiative transition and then progressively mixes back into the bulk as the optical depth increases.
Although the detailed mechanism of this remixed conversion remains to be specified, this motivates a multi-population description in which the local distribution is represented as the superposition of a bulk and a cooled component.
The remaining task is to determine how their relative weights vary with optical depth and how this evolving partition reconstructs the mean entropy profile.

\section{Two-population closure for the surface transition}
\label{sec:multi_population}

We now turn this phenomenology into a minimal extension of the phase-space formalism.
The aim is to retain the maximum-entropy bulk described in Sect.~\ref{sec:1D_closure}, while allowing for the radiatively cooled contribution identified in the surface layers.
We first formulate the corresponding decomposition for a general set of populations, before specialising it to the bulk--plume partition suggested by simulations.

\subsection{Multi-population formalism}

We begin by generalising the PSA formalism of Paper~I to a phase-space density partitioned into \(N\) populations, and by examining what this implies for the transport of these sub-distributions and for the decomposition of the mean fields.

\subsubsection{Phase-space continuity with several populations}
\label{sec:multipop_continuity}

We let the total phase-space density be partitioned as
\begin{equation}
        \rho = \sum_i \rho_i,
\end{equation}
where \(\rho_i(t,\vec r,\vec u)\) denotes the phase-space mass density associated with population \(i\).
It is convenient to introduce the corresponding local phase-space fractions
\begin{equation}
        X_i \equiv \frac{\rho_i}{\rho},
        \qquad
        \sum_i X_i = 1.
\end{equation}

While the total phase-space density satisfies the continuity equation~\eqref{eq:phase_space_continuity}, this need not be the case for each sub-distribution separately, since the relative importance of the different populations may vary across phase space.
To represent this, we introduce exchange terms \(\Pi_{ij}\) describing local mass conversion from population \(j\) to population \(i\), per unit total mass.
The population densities then satisfy
\begin{equation}
        \partial_t \rho_i
        +
        \grad_{r} \cdot(\rho_i \vec u)
        +
        \grad_{u} \cdot(\rho_i \vec\gamma)
        =
        \rho \sum_j \Pi_{ij}.
\end{equation}
By construction, the exchange matrix is antisymmetric,
\begin{equation}
        \Pi_{ij} = -\Pi_{ji},
\end{equation}
so summation over \(i\) recovers the continuity equation of the total phase-space density \eqref{eq:phase_space_continuity}.
Using this global relation, the evolution of the local fractions may be written in the compact form
\begin{equation}
        \label{eq:phase_space_fraction}
        \Dt X_i = \sum_j \Pi_{ij},
        \qquad
        \Dt \equiv \partial_t + \vec u \cdot \grad_{r} + \vec\gamma \cdot \grad_{u}.
\end{equation}
Knowing how the phase-space weights associated with each population evolve therefore reduces to specifying the rates of mass exchange between them.

\subsubsection{Decomposition of the mean fields}
\label{sec:multipop_mean_decomposition}

Exactly as for the total distribution, one may define for each population the mean density,
\begin{equation}
        \bar\rho_i(\vec r,t) = \int \rho_i(\vec r,\vec u,t)\,d\vec u,
        \qquad
        \bar\rho = \sum_i \bar\rho_i,
\end{equation}
and the associated conditional velocity distribution,
\begin{equation}
        f_i(\vec u \mid \vec r,t)
        \equiv
        \frac{\rho_i(\vec r,\vec u,t)}{\bar\rho_i(\vec r,t)},
        \qquad
        \langle f_i \rangle = 1.
\end{equation}
The corresponding mean population fractions are simply
\begin{equation}
        \tilde X_i
        \equiv
        \langle f X_i \rangle
        =
        \frac{\bar\rho_i}{\bar\rho},
        \qquad
        \sum_i \tilde X_i = 1.
\end{equation}
The global conditional distribution may then be decomposed as
\begin{equation}
        f(\vec u \mid \vec r,t)
        =
        \sum_i \tilde X_i(\vec r,t)\, f_i(\vec u \mid \vec r,t),
\end{equation}
so that the Favre average of any specific quantity \(x(\vec r,\vec u,t)\) may be written as
\begin{equation}
        \tilde x = \langle f x \rangle = \sum_i \tilde X_i \, \langle f_i x \rangle.
\end{equation}

This rewriting is more than a formal decomposition.
It replaces the description of a single local distribution by a partition over simpler sub-distributions, in the hope that these elementary contributions may themselves admit closed forms.
The mean stratification is then reconstructed from the conditional fields of the different populations together with their relative weights.

To see what this changes in practice, we consider the convective flux,
\begin{equation}
        Q_c = \langle \rho H'' u \rangle,
\end{equation}
where \(H = h + k + \Phi\) denotes the Hamiltonian.
Using the decomposition above, one is naturally led to separate two distinct contributions,
\begin{equation}
        Q_c = \bar\rho \sum_i \tilde X_i \left( H_i u_i + \phi_i \right),
\end{equation}
with
\begin{ceqn}    
        \[
                u_i = \langle f_i u \rangle,
                \qquad
                H_i = \langle f_i H \rangle,
                \qquad
                \phi_i \equiv \langle f_i H'' u'' \rangle,
        \]
\end{ceqn}
where the double primes denote here fluctuations with respect to the mean of population \(i\), rather than the global mean.
Indeed, the condition of vanishing total mass flux only implies
\begin{equation}
        \sum_i \tilde X_i u_i = 0,
\end{equation}
so the individual population velocities need not vanish separately in the present framework.

\subsection{Two-population closure}
\label{sec:two_population_closure}

We now specialise the general formalism above to the minimal case suggested by Sect.~\ref{sec:phenomenology_3D}, namely a decomposition into a convective bulk population \((+)\) and a plume population \((-)\).
The \((+)\) population is therefore identified with the broad maximum-entropy background already described in Sect.~\ref{sec:1D_closure}, while the \((-)\) population is introduced to represent the plume component singled out by the surface transition and visible on the low-entropy side of Fig.~\ref{fig:entropy_slices}.

The mean entropy assigned to the plume population follows from the radiative organisation described in Sect.~\ref{sec:phenomenology_3D}.
The cooled branch follows the radiative entropy branch \(s_0\) through the surface transition, then preserves its entropy deficit as it descends and progressively mixes back into the bulk.
To capture this behaviour at leading order, we prescribed
\begin{equation}
        \label{eq:entropy_plume}
        \frac{d s_-}{d\tau} = \min \left(\frac{d s_0}{d\tau},\,0\right),
\end{equation}
where \(s_0(\tau)\) denotes the radiative entropy branch constructed from the grey Eddington reference profile.
Thus the plume population follows the radiatively constrained branch where cooling is active, and descends adiabatically once it has entered the convective region.

The bulk population is, by definition, the part of the local distribution that remains organised by the exponential maximum-entropy closure derived in Sect.~\ref{sec:1D_closure}.
Its parameters are therefore the natural generalisations of those introduced there.
The entropy contrast and maximum upper velocity are now measured with respect to the mean bulk properties:
\begin{equation}
        s_+ \equiv \langle f_+ s \rangle \neq \tilde s,
        \qquad
        u_+ \equiv \langle f_+ u \rangle \neq \tilde u = 0,
\end{equation}
so the replacements
\begin{ceqn}
        \[
                S-\tilde s \;\longrightarrow\; S-s_+,
                \qquad
                U \;\longrightarrow\; U-u_+
        \]
\end{ceqn}
must be made.
Within this population, the exponential closure and the associated convective optimum are therefore retained in the same form as in Sect.~\ref{sec:1D_closure}.
In what follows, we continue to denote by \(\phi\) the turbulent flux associated with this bulk component.

The new point concerns the way the convective flux is partitioned once the two populations coexist.
Because the plume distribution is comparatively much more symmetric and homogeneous in its thermodynamic properties than the bulk distribution, its internal turbulent transport is expected to remain weak.
At leading order, we therefore neglect the intra-population plume flux \(\phi_- = \langle f_- H'' u'' \rangle\).
Using the vanishing of the total mass flux, the convective flux balance then reduces to
\begin{equation}
        \label{eq:flux_balance_two_populations}
        Q_r + \bar{\rho} (\tilde X_+ \phi + \tilde X_- \phi_{\rm rel}) = Q,
\end{equation}
where the inter-population contribution, \(     \phi_{\rm rel} = (H_- - H_+) u_-\), may become important near the surface.

The consequence of this decomposition is simple.
Once plumes are present, the transport of total energy no longer fixes the turbulent bulk flux \(\phi\) on its own.
In the mono-population case, the same balance \eqref{eq:flux_balance} directly imposed \(\bar \rho \phi = Q - Q_r\),  whereas Eq.~\eqref{eq:flux_balance_two_populations} now constrains only the weighted sum of the bulk and relative contributions.
The passage to two populations therefore leaves open a new question: now that the realised convective flux \(Q-Q_r\) and the turbulent bulk flux \(\bar\rho\phi\) are distinct quantities, what value of \(\phi\) should actually enter the convective optimum \eqref{eq:optimum_section2} of the exponential bulk?

\subsection{A generalised convective optimum}
\label{sec:interpretation_optimum}

The distinction introduced by Eq.~\eqref{eq:flux_balance_two_populations} is central to the two-population closure.
The first, \(\bar\rho\phi\), is the flux scale entering the bulk optimum~\eqref{eq:optimum_section2} and measures the transport associated with the internal organisation of the maximum-entropy bulk.
The second, \(Q-Q_r\), is the stationary convective flux resulting from the full two-population balance~\eqref{eq:flux_balance_two_populations}, after radiative transfer, plume production, and relative transport have all contributed.

These two quantities play different roles.
Using \(Q-Q_r\) as the flux entering the bulk optimum would implicitly impose a sequential picture in which radiative transport first selects the residual convective flux and the bulk then reorganises in response to it.
This ordering is not appropriate in the convective region, where the bulk reorganises on a timescale at least as short as the radiative one, and much shorter in the deep unstable layers.
We therefore adopt the hypothesis that the local bulk state is set by the total imposed stellar flux, while the partition between radiative and convective transport emerges from the stationary balance.
This corresponds to an internal bulk optimum of the form
\begin{equation}
        \bar\rho\phi = Q.
\end{equation}
Radiative transfer then acts by limiting the fraction of the local distribution that remains in the bulk state:
\begin{enumerate}[(i)]
        \item In thermally stable layers, this limitation is complete: the bulk cannot form at all, so 
        \begin{ceqn}
                \[
                        \tilde X_+ \simeq 0,
                \]
        \end{ceqn}
        and the flux is carried entirely by radiation.
        \item In deeply unstable layers, by contrast, radiative losses are too slow to disrupt the exponential organisation of the flow, the bulk occupies essentially the whole distribution, and one recovers the pure-bulk regime with 
        \begin{ceqn}
                \[
                        \tilde X_+ \simeq 1.
                \]
        \end{ceqn}
        \item The surface layers correspond to the intermediate case: the bulk is still represented by the same internal optimum, but it can no longer occupy the whole local distribution.
        The hottest upflows are progressively peeled away by radiative cooling and converted into plumes so that their fraction decreases through the surface transition and
        \begin{ceqn}
                \[
                        0 < \tilde X_+ < 1.
                \]
        \end{ceqn}
\end{enumerate}
The net convective flux may therefore be smaller than \(Q\) simply because the bulk is no longer realised throughout the whole distribution.

\begin{figure*}[!ht]
    \centering
    \includegraphics[width=\textwidth]{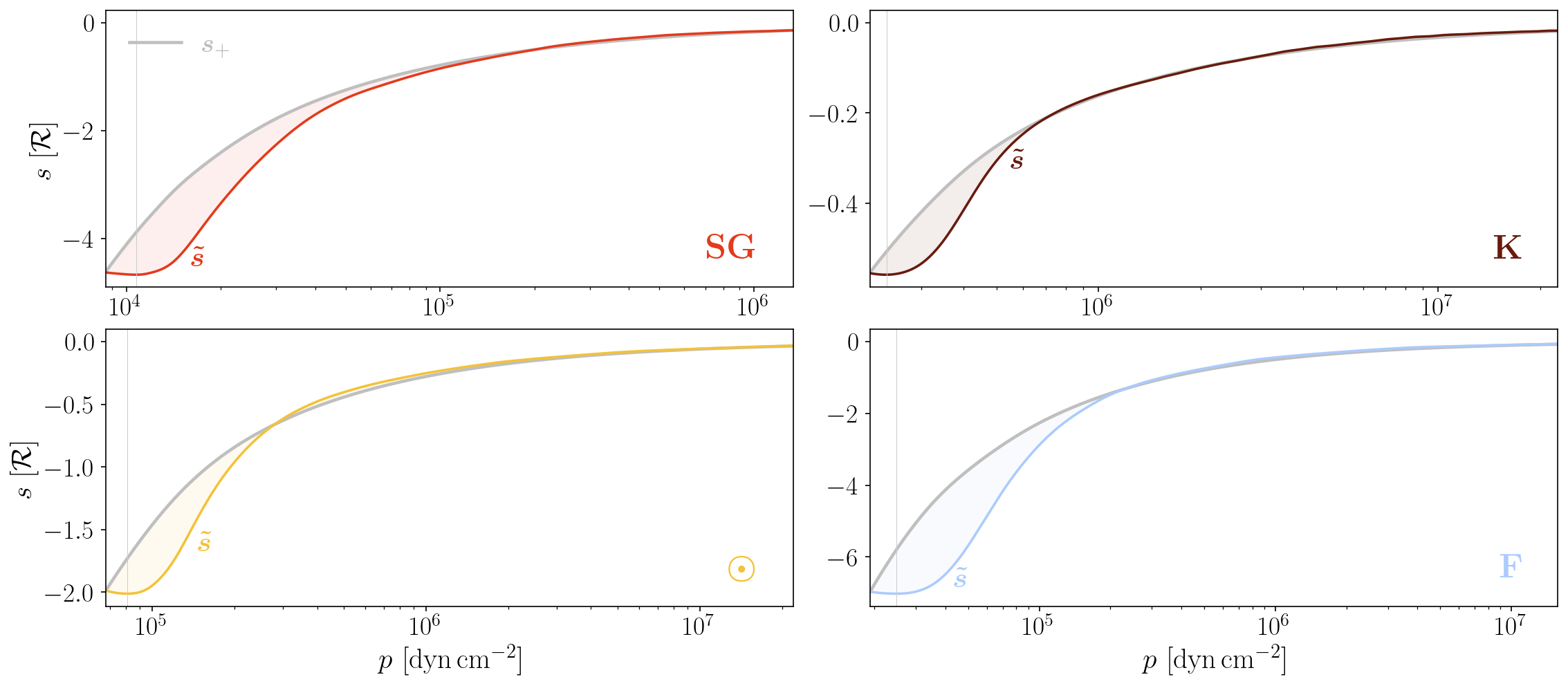}
        \caption{
        Comparison between the entropy profile \(s_+\) associated with the optimal bulk state, Eq.~\eqref{eq:entropy_bulk}, and the simulated mean entropy profile \(\tilde s\) for the four reference atmospheres, shown as functions of gas pressure.
        In each panel, the grey curve denotes \(s_+\), while the coloured curve denotes \(\tilde s\).
        The panels correspond respectively to the subgiant (SG; red, top left), the K dwarf (K; brown, top right), the solar model (\(\odot\); gold, bottom left), and the F dwarf (F; light blue, bottom right).
        As in Figs.~\ref{fig:hist_entropy}-\ref{fig:hist_entropy_cond}, the vertical line marks the approximate location of the photosphere.
        }
    \label{fig:entropy_optimum}
\end{figure*}

This unified reading of the transport regimes leads directly to a structural prediction.
If the pure bulk state, with \(\tilde X_+=1\) and \(\bar\rho\phi=Q\), represents the least constrained local organisation of the convective flow, then mixing this bulk with radiatively cooled material can only lower the resulting mean entropy relative to the pure bulk profile.
Combining the bulk optimum~\eqref{eq:optimum_section2} with \(\bar\rho\phi=Q\) gives
\begin{equation}
        \label{eq:entropy_bulk}
        s_+ = S - \frac{3}{\tilde T}\left(\frac{Q}{2\bar\rho}\right)^{2/3}.
\end{equation}
One therefore expects, throughout the convective region where this bulk reference remains meaningful,
\begin{ceqn}
        \[
                \tilde s \leq s_+.
        \]
\end{ceqn}
In other words, once the bulk is only partially realised, the mean entropy profile must fall below the one associated with the optimal bulk alone.
The best entropy stratification effectively accessible to convection is therefore not the unattained limit \(s=S\), but the entropy profile \(s_+\) associated with the optimal bulk state.
The important point is that the reference profile \(s_+\) itself is fixed by the bulk closure.
The resulting prediction can therefore be tested directly against the simulated mean entropy, without identifying or decomposing the flow into bulk and cooled populations.

Figure~\ref{fig:entropy_optimum} shows that this behaviour is found in all four reference atmospheres introduced in Sect.~\ref{sec:3D_models}.
Read from right to left, from the deeper layers towards the surface, the same organisation appears in each panel.
Deep in the convective region, the simulated mean entropy closely follows the optimal bulk profile \(s_+\), as expected when the flow is dominated by the bulk dynamics.
Approaching the surface transition, the two profiles separate systematically: the realised mean entropy bends below \(s_+\), with the largest offset occurring near the entropy minimum marked by the vertical line.
Within the two-population interpretation developed below, this is precisely the region where radiatively cooled material becomes an important component of the flow.

Seen in this way, Fig.~\ref{fig:entropy_optimum} shows how remarkably well the simple closed-form prediction~\eqref{eq:entropy_bulk} captures the mean entropy stratification over most of the convective region, and that this remains true across all four simulations despite their very different global parameters.
This agreement is all the more striking because the expression for the entropy deficit, \(S - s_+\), is both compact and transparent: it is set by the imposed flux per unit mass, \(Q/\bar\rho\), and the local mean temperature \(\tilde T\).
No adjustment to the simulated entropy profile enters this estimate.

\subsection{Determination of the plume fraction}
\label{sec:plume_fraction}

Now that the mean entropies of the two populations, \(s_+\) and \(s_-\), have been specified, the reconstruction of the mean thermal stratification reduces to the determination of their relative fractions.
Indeed, the two-population decomposition gives
\begin{equation}
        \tilde s = \tilde X_+ s_+ + \tilde X_- s_-,
        \qquad
        \tilde X_+ + \tilde X_- = 1,
\end{equation}
so the whole problem may be reformulated in terms of the single unknown \(\tilde X_-\).
Once this quantity is known, the mean entropy profile follows immediately.

The point of view adopted here is to follow the radiatively constrained material as it penetrates the convective region.
In the optically thin radiative layers, where the thermal structure is set by radiative adjustment, this component represents essentially the whole fluid, so that one has \(\tilde X_-\simeq 1\).
Below the surface transition, the same material is carried downwards by the plumes and progressively loses its identity as it remixes with the surrounding bulk.
The problem considered here is therefore to describe the evolution of the plume fraction \(X_-\) during that descent.

Specialising the general population balance~\eqref{eq:phase_space_fraction} to the two-population case and denoting the local conversion rate from plume material into the bulk by \(\Pi_{+-}>0\), we obtain
\begin{equation}
        \Dt X_- = -\Pi_{+-}.
\end{equation}
As suggested by the entropy histograms of Fig.~\ref{fig:entropy_slices}, the distinction between the two populations is both thermodynamic and statistical.
The bulk population is represented by the broad maximum-entropy distribution, whereas recently cooled material retains a more constrained organisation in phase space.
Equation~\eqref{eq:statistical_entropy_material} identifies \(\{s,T\}\) as the local statistical-entropy production rate in an expansive region.
We therefore used it as the dynamical rate in the minimal mixing law
\begin{equation}
        \Pi_{+-} = \{s,T\}\,X_+X_-.
\end{equation}
The factor \(X_+X_-\) ensures that conversion vanishes when either population is absent.

Passing to the mean plume fraction and neglecting, at leading order, the variation of the plume mass flux \(\bar\rho u_-\) along the descent, one obtains
\begin{equation}
        D_t \tilde X_- = -\,\langle f \Pi_{+-} \rangle,
\end{equation}
where \(D_t\) denotes the material derivative along the mean plume motion.
Because \(f X_i = f_i \tilde X_i\),
\begin{equation}
        X_+ X_- = \frac{f_+f_-}{f^2} \, \tilde X_+ \tilde X_-,
\end{equation}
and therefore
\begin{equation}
        D_t \tilde X_- = -\Lambda\, \tilde X_+ \tilde X_-,
\end{equation}
with
\begin{equation}
        \Lambda \equiv \langle g \{s,T\}\rangle,
        \qquad
        g \equiv \frac{f_+f_-}{f}.
\end{equation}
Here \(g\) measures the phase-space overlap entering the effective conversion rate.
We define the cumulative conversion measure
\begin{equation}
        d\mu = \Lambda\,dt,
\end{equation}
which increases along the plume descent as long as \(\Lambda\geq0\).
The evolution equation then becomes
\begin{equation}
        d\tilde X_-
        =
        -\tilde X_-(1-\tilde X_-)\,d\mu.
\end{equation}
Its solution is logistic,
\begin{equation}
        \tilde X_-(\mu)
        =
        \frac{e^{-\mu}}{1+e^{-\mu}},
        \qquad
        \tilde X_+(\mu)
        =
        \frac{1}{1+e^{-\mu}},
\end{equation}
with \(\mu\) defined up to an integration constant.

The remaining problem is to relate the conversion measure \(\mu\) to the one-dimensional stratification.
A directly usable expression does not follow from \(\Lambda\) alone, since the phase-space overlap entering its definition is not known from the mean structure.
We therefore sought a monotonic stratification coordinate over which the transition between the two populations is approximately universal.
The pressure coordinate does not provide such a description for instance: Fig.~\ref{fig:entropy_optimum} shows that the transition spans substantially different ranges of \(\ln p\) among the four atmospheres.
Empirically, a much more uniform behaviour is obtained by taking
\begin{equation}
        d\mu \simeq d\ln\tau,
\end{equation}
which gives
\begin{equation}
        \mu = \ln \left(\frac{\tau}{\tau_0}\right).
\end{equation}
The integration constant \(\tau_0\) then has a simple interpretation: it is the optical depth at which the two populations have equal weights.
With this choice, the plume and bulk fractions become
\begin{equation}
        \tilde X_-(\tau)=\frac{\tau_0}{\tau+\tau_0},
        \qquad
        \tilde X_+(\tau)=\frac{\tau}{\tau+\tau_0}.
\end{equation}

The remaining unknown is \(\tau_0\).
In the present picture, radiative cooling progressively removes the high-entropy part of the bulk distribution as the radiative branch \(s_0(\tau)\) moves through it.
We identify the point of equal population weights, \(\tilde X_+=\tilde X_-=1/2\), with the stage at which this limiting branch has reached the median entropy of the pure bulk distribution.
For the maximum-entropy closure of Sect.~\ref{sec:1D_closure}, the entropy deficit \(S-s\) is exponentially distributed with mean \(\delta s\) and therefore has a median of \(\ln2\,\delta s\).
The transition optical depth is consequently fixed by
\begin{equation}
        \label{eq:tau0_median_criterion}
        s_0(\tau_0)
        =
        S-\ln2\,\delta s(\tau_0).
\end{equation}
This closes the two-population model without introducing any additional free parameter.
Once \(\tau_0\) is known, the mean entropy profile follows directly:
\begin{equation}
        \label{eq:entropy_2pop}
        \tilde s
        =
        \frac{\tau s_+ + \tau_0 s_-}{\tau+\tau_0}.
\end{equation}

\section{Two-population closure: Comparison with three-dimensional simulations}
\label{sec:comparison_3D}

\begin{figure*}[!ht]
        \centering
        \includegraphics[width=\textwidth]{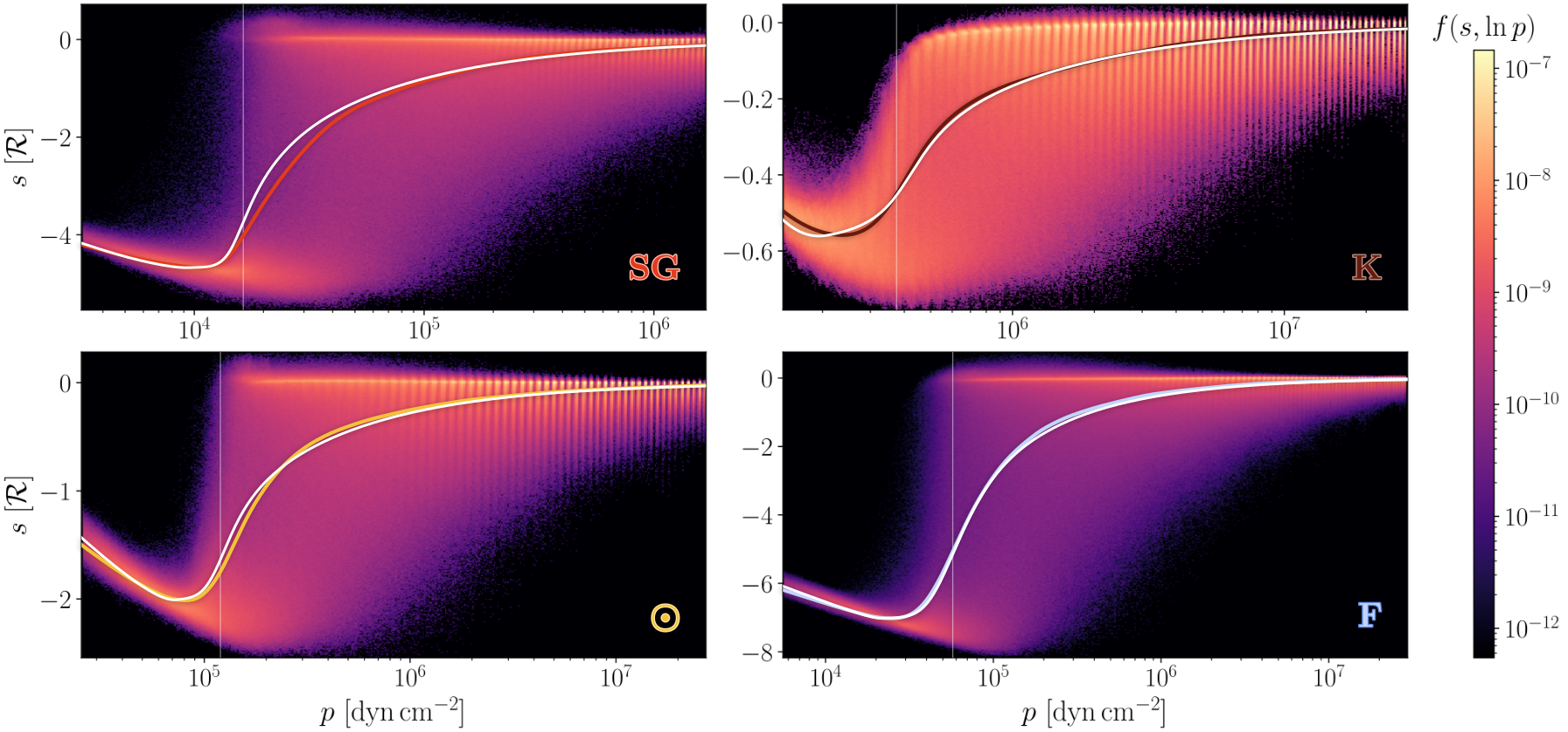}
        \caption{
        Mass-weighted entropy distributions in the \((p,s)\) plane for the four reference simulations.
        The colour scale shows the layer-normalised sampling of \(f(s,\ln p)\).
        The coloured curves show the simulated Favre-mean entropy profiles \(\tilde{s}_{\rm 3D}\), and the white curves show the two-population reconstruction \(s_{\rm 2pop}=\tilde X_+s_+ + \tilde X_-s_-\).
        The thin vertical lines mark the values of \(\tau_0\) defined by Eq.~\eqref{eq:tau0_median_criterion}.
        }
        \label{fig:two_pop_entropy}
\end{figure*}

Figure~\ref{fig:two_pop_entropy} compares the two-population reconstruction with the entropy distributions of the four reference simulations.
The background histograms show the same mass-weighted quantity as in the previous statistical diagnostics, now for all four models introduced in Sect.~\ref{sec:3D_models}.
They reveal both a common organisation and clear model-to-model differences.
In all cases, the deep layers remain dominated by a high-entropy bulk branch, while the surface transition is associated with the development of a lower-entropy contribution.
The visual separation between these branches becomes more pronounced as the entropy jump increases.
In the F dwarf, for instance, the distribution is close to two sharply populated branches connected by a more weakly populated transition region, whereas the K dwarf shows a smoother evolution between the deep bulk and the cooled component.
The solar model and the subgiant occupy intermediate but distinct regimes, illustrating that the transition morphology varies substantially across the range of effective temperatures and gravities considered here.

The coloured curves show the simulated Favre-mean entropy profiles, while the white curves show the reconstruction obtained from the closure derived in Sect.~\ref{sec:multi_population}.
No direct fit to \(\tilde s_{\rm 3D}\) is performed.
The two entropy branches \(s_+\) and \(s_-\) are fixed respectively by the bulk closure of Eq.~\eqref{eq:entropy_bulk} and by the radiatively cooled branch of Eq.~\eqref{eq:entropy_plume}.
The mean profile is then obtained from Eq.~\eqref{eq:entropy_2pop}, with \(\tau_0\) determined by the exponential-median criterion of Eq.~\eqref{eq:tau0_median_criterion}.

The agreement is remarkably close across all four atmospheres.
In the deep layers, the reconstruction remains locked to the bulk branch, as expected when \(\tilde X_+\simeq 1\).
Across the surface transition, the progressive increase of \(\tilde X_-\) lowers the mean entropy and reproduces the departure of the simulated profile from the pure-bulk prediction highlighted in Fig.~\ref{fig:entropy_optimum}.
The same behaviour is recovered from the cool high-gravity K dwarf to the hot F dwarf and the low-gravity subgiant, despite the substantially different entropy distributions shown in the background.
In this sense, Fig.~\ref{fig:two_pop_entropy} provides a non-trivial test of the two-population closure over a broad range of envelope conditions, without any direct calibration to the simulated mean entropy profiles.

The quantitative level of agreement, as well as the residual differences, should nevertheless be interpreted with some care.
The closure itself relies on several simplifying assumptions, some of which are only approximately realised in the simulations.
Conversely, the simulations do not provide an exact physical reference: in the layers considered here, their detailed stratification may be influenced in particular by the lower boundary conditions and by the numerical treatment of viscous dissipation, as well as by the finite computational domain and the adopted radiative-transfer treatment.
Close agreement should therefore not be taken as validation of each individual assumption entering the closure, just as local discrepancies cannot be assigned unambiguously to a failure of the closure alone.
The comparison is instead most informative through the systematic character of the agreement across the four contrasting atmospheres.
We return to these broader questions in Sect.~\ref{sec:discussion}.

\section{Discussion}
\label{sec:discussion}

\subsection{The upper-edge structure of the simulated bulk distribution}
\label{sec:discussion_upper_edge}

The local bulk distribution discussed in Sect.~\ref{sec:phenomenology_3D} displays the two main features anticipated by the maximum-entropy construction: a low-entropy tail close to an exponential form, and a sharp upper edge near the entropy level \(S\).
Beyond these two features, however, the detailed shape of the distribution close to its upper edge is also informative.
Instead of reaching this edge with the smooth exponential weight expected from the ideal maximum-entropy distribution, the simulated distribution is more strongly concentrated near \(S\), as visible in Fig.~\ref{fig:bulk_entropy_distribution}.
Near the surface, some layers also show a small population above the nominal bound, as seen in Fig.~\ref{fig:two_pop_entropy}.

This excess concentration can be understood by looking at the entropy sources present in the simulations.
Schematically, the material entropy equation contains two types of source terms: radiative heating or cooling and viscous or effective numerical dissipation.
In the optically thick part of the transition, the radiative term acts primarily as a smoothing mechanism: it reduces thermal contrasts through diffusion and is therefore not the most natural source of rare fluid elements with entropies above the bulk reference.
By contrast, viscous heating is positive definite and becomes important where the flow is strongly braked.
The high-entropy excursions visible near and slightly above \(S\) appear precisely in the sub-surface region where the ascending motions are decelerated most strongly.
This makes effective viscous or hyper-viscous heating a plausible origin for these exceptional elements.
It may also have a less visible effect, by accumulating material close to the maximum entropy level without necessarily producing a clearly separated population above \(S\).

A second contribution is likely associated with the lower boundary condition.
The local boxes do not include the asymptotic deep adiabat itself, so the high-entropy reservoir feeding the convective flow has to be represented at the bottom of the computational domain.
In the simulations used here, the bottom entropy is kept fixed and the fixed-entropy condition is enforced through isentropic inflows, while outflows are left unchanged.
This prescription continuously injects upward-moving material with a preferred entropy, while the phase-space expansion associated with \(\{s,T\}>0\) progressively dilutes this boundary imprint as the material moves away from the lower boundary.
A surplus of material near \(s=S\) at the bottom of the box is therefore expected, and this is indeed what is seen in the deepest layers, where the distribution contains an excess of samples at, or very close to, the imposed entropy level.
Part of this imprint can persist over the finite depth of the simulated domain.

\subsection{Interpretation of the conversion coordinate}
\label{sec:discussion_conversion_coordinate}

The two-population closure relies on the conversion measure \(\mu\), introduced through
\[
        d\mu = \Lambda\,dt,
        \qquad
        \Lambda = \langle g\{s,T\}\rangle.
\]
As discussed in Sect.~\ref{sec:plume_fraction}, the effective rate \(\Lambda\) combines two ingredients: the phase-space overlap between the plume and bulk distributions, measured by \(g=f_+f_-/f\), and the local dynamical rate of remixed conversion, measured by the phase-space divergence \(\{s,T\}\).
The overlap factor vanishes when the two populations occupy disjoint regions of phase space, so that conversion requires their simultaneous local presence.
The quantity \(\mu\) therefore measures the cumulative opportunity for conversion along the plume descent.

Appendix~\ref{app:baroclinic_limit} provides a possible local interpretation of the dynamical factor entering \(\Lambda\).
For a low-Mach, high-Reynolds plume--bulk interface in which the vorticity balance is dominated by baroclinic production, the phase-space divergence may be related to the growth of the projected baroclinic vorticity, \(\omega\), by
\begin{equation}
        D_t\ln\omega^2 \simeq \{s,T\}.
\end{equation}
Because the relevant vorticity field can be regarded as independent of the velocity coordinate entering the local phase-space average, the conversion measure may then be read schematically as
\begin{equation}
        d\mu
        \simeq
        \langle g\rangle\,d\ln\omega^2,
        \qquad
        0 \leq \langle g\rangle \leq 1.
\end{equation}
Under this interpretation, the plume population loses its identity through the cumulative baroclinic stirring generated at the interface between the cooled material and the surrounding bulk.
The overlap factor \(\langle g\rangle\) modulates this process by measuring how much of the two populations share the same local region of phase space.
In this picture, remixed conversion is favoured when cooled and bulk material remain in contact while having sufficiently similar velocities to interact over a finite time.
The baroclinic vorticity generated by the contrast in specific volume then provides a natural mechanism for maintaining such contact, by producing horizontal vortical motions that keep neighbouring elements of the two populations in a common mixing region.
The conversion measure \(\mu\) may thus be viewed as a cumulative measure of this mixing process.

This interpretation does not, however, provide a directly usable one-dimensional closure.
Neither the phase-space overlap nor the accumulated vorticity growth along a plume trajectory is readily predicted from the mean stratification.
The identification \(d\mu \simeq d\ln\tau\) should therefore be regarded as a separate phenomenological assumption.

The use of optical depth as an effective conversion coordinate is motivated by the fact that the conversion occurs across the radiative transition region, where the radiative environment changes rapidly over a relatively narrow geometrical region.
Using \(\ln\tau\) further assumes that comparable relative changes in optical depth correspond to comparable increments of conversion.
The systematic agreement obtained in Fig.~\ref{fig:two_pop_entropy} suggests that this minimal prescription is sufficient for the present set of atmospheres, while a more direct connection between \(\mu\) and the underlying mixing dynamics remains to be established.

\subsection{Towards one-dimensional stellar models}
\label{sec:discussion_1d_models}

\subsubsection{From layer stability to population fractions}

A first consequence of the two-population closure is that the radiative--convective transition is no longer described primarily as a boundary between layers of different nature.
In standard one-dimensional modelling, a local stability criterion is usually used to decide whether a given layer should be treated as radiative or convective.
The convective contribution is then controlled by an almost binary indicator: in a thermally stable layer the convective closure is inactive, whereas in an unstable layer it is switched on.

The present framework suggests a continuous generalisation of this construction.
The relevant question is not only whether the mean stratification is unstable, but what fraction of the local distribution actually belongs to the bulk population.
This role is played by the realised bulk fraction \(\tilde X_+\).
In the thermally stable limit, \(\tilde X_+\simeq 0\), and the stratification is entirely controlled by the radiative branch.
In the deeply unstable limit, \(\tilde X_+\simeq 1\), and the exponential bulk occupies essentially the whole distribution.
Between these limits, both populations coexist, so that the mean stratification reflects their relative weights rather than the assignment of the layer to a single regime.

This suggests a deeper change in the way one-dimensional stellar structures are organised.
Rather than being assembled from distinct radiative and convective regions, the model could in principle be treated as a single continuous domain governed by the same set of mean equations.
The usual radiative, convective, and transitional regimes would then appear as limiting cases of the same population decomposition, with their local character determined by the value of \(\tilde X_+\).

\subsubsection{A self-consistent stellar construction}

The status of the deep entropy level \(S\) depends on the problem under consideration.
In the comparison with simulations, there is no ambiguity.
The deep entropy level is read directly from the three-dimensional model,
\begin{equation}
        S = S_{\rm 3D},
\end{equation}
and is not adjusted to reproduce the simulated mean entropy profile.
Once \(S\) and the mean thermodynamic stratification are given, the radiative branch \(s_0\), the cooled branch \(s_-\), the bulk branch \(s_+\), and the transition depth \(\tau_0\) are specified, respectively, by Eqs.~\eqref{eq:entropy_radiative_branch}, \eqref{eq:entropy_plume}, \eqref{eq:entropy_bulk}, and \eqref{eq:tau0_median_criterion}.
The reconstructed mean entropy profile then follows from Eq.~\eqref{eq:entropy_2pop}.

The situation is different when the closure is embedded in a complete stellar model, in which case the deep entropy level \(S\) is determined by the constraints imposed by the stellar interior.
To see this, consider first a decomposition of the star into an inner radiative region and an outer convective envelope.
For fixed composition and microphysics, prescribing the mass \(M_{\rm rad}\) and luminosity \(L_{\rm rad}\) of the radiative interior provides the global constraints required to determine its regular structure, for instance by adjusting the central thermodynamic state.
The thermodynamic state reached at the upper boundary of the radiative interior then determines the entropy of the deep convective adiabat.
The overlying envelope can be subsequently integrated outwards using the present closure, with continuity of the thermodynamic stratification at the matching point.
In this construction, \(S\) is therefore an output of the interior--envelope matching rather than an adjustable convective-efficiency parameter.

More generally, stellar models are specified by their total mass and luminosity, \(M_*\) and \(L_*\), rather than by \(M_{\rm rad}\) and \(L_{\rm rad}\).
In this case, \(M_{\rm rad}\), \(L_{\rm rad}\), and \(S\) are not prescribed independently.
For a given prescription of the radiative--convective transition, each trial central thermodynamic state, \(\vec x_c\), determines through integration of the stellar interior a corresponding matching state, \(\vec y_{\rm rad}\), of which \(M_{\rm rad}\), \(L_{\rm rad}\), and \(S\) are components:
\begin{ceqn}
        \[
                \vec x_c \longmapsto \vec y_{\rm rad}(\vec x_c).
        \]
\end{ceqn}
This state then determines the response of the overlying convective envelope following the procedure described above.
The coupled interior--envelope construction may therefore be represented schematically as
\begin{ceqn}
        \[
                \vec x_c
                \longmapsto
                \vec y_{\rm rad}(\vec x_c)
                \longmapsto
                \Big\{
                        M(\vec x_c),
                        L(\vec x_c),
                        R(\vec x_c)
                \Big\}.
        \]
\end{ceqn}
A self-consistent model may then be obtained by iterating over the central thermodynamic state until the global mass and luminosity constraints are simultaneously satisfied:
\begin{equation}
        M(\vec x_c)=M_*,
        \qquad
        L(\vec x_c)=L_*.
\end{equation}
The deep entropy level \(S\) is therefore an intermediate quantity selected by the coupled interior--envelope construction rather than an independent shooting parameter, while the stellar radius \(R\) is determined as part of the same global solution.

The feedback through the stellar geometry is retained explicitly through the quantities entering the envelope closure:
\begin{equation}
    Q(r) = \frac{L(r)}{4\pi r^2},
    \qquad
    g(r) = \frac{Gm(r)}{r^2},
\end{equation}
so that changes in the entropy stratification feed back on the stellar radius and therefore on the local flux and gravity.
This has an important consequence: rather than adjusting \(S\) so as to match a target radius \(R_*\) at a fixed mass and luminosity, as effectively done through the calibration of the convective efficiency in MLT-type models, the present closure leaves no additional convective parameter with which to adjust the radius, which instead emerges as a prediction \(R(M_*,L_*)\) of the self-consistent stellar model.

An isolated envelope constitutes a special limiting problem.
In a plane-parallel calculation with externally prescribed \(Q\) and \(g\), the geometrical feedback described above is absent.
The closure then defines a family of envelope structures parameterised by the deep entropy level \(S\), which must be supplied as an additional boundary condition.
Such an isolated envelope therefore requires \(Q\), \(g\), and \(S\) to be specified independently.

\section{Conclusion}
\label{sec:conclusion}

In this paper, we have developed a phase-space closure for the outer convective layers of cool stars with the aim of connecting the statistical structure of the flow to one-dimensional stellar modelling.
The starting point is the maximum-entropy description of the convective bulk.
Under the assumption that the realised bulk states are bounded by a deep entropy level \(S\), the local velocity distribution takes a one-sided exponential form.
This provides a simple relation between the imposed stellar flux, the entropy deficit of the bulk, and the mean stratification associated with the optimal convective state.

The comparison with three-dimensional radiation-hydrodynamics simulations shows that this construction captures a robust feature of the deep convective region.
The local entropy distribution is sharply bounded on its high-entropy side, and it develops a low-entropy tail close to the predicted exponential form.
As the surface is approached, however, the distribution changes character.
Radiative cooling produces a distinct low-entropy component, while the high-entropy branch remains tied to the radiative constraint imposed by the atmosphere.

Motivated by this behaviour, we introduced a two-population closure in which the bulk population is described by the optimal maximum-entropy state, while the cooled population follows the radiative entropy branch through the surface transition and then descends adiabatically.
The mean entropy profile is reconstructed from the relative weights of these two branches.
In this description, the realised bulk fraction \(\tilde X_+\) controls the local character of the mean stratification: radiative, transitional, and bulk-convective behaviours correspond to different limits of the same population decomposition, rather than to separate layer-wise prescriptions.
The resulting model reproduces the mean entropy stratification of the four reference simulations over most of the displayed range without adjusting the entropy profile itself.
The remaining phenomenological ingredient is the conversion coordinate, for which \(d\mu\simeq d\ln\tau\) provides a simple empirical closure. Establishing a more direct connection with phase-space overlap, baroclinic stirring, and radiative transport remains an important theoretical question.

The natural next step is to embed this closure in self-consistent one-dimensional stellar models.
At fixed stellar parameters and microphysics, the coupled interior--envelope solution selects the deep entropy level \(S\), while the closure determines the surface entropy connection from the population structure of the flow.
In this setting, both the surface entropy jump and the resulting stellar radius emerge without recourse to a calibrated convective-efficiency parameter.
This makes well-characterised stars especially valuable tests of the approach.
The Sun provides the natural reference case, while \(\alpha\) Cen AB offers a particularly stringent binary test: its two components share a common age and initial composition, their radii are constrained interferometrically, and both show solar-like oscillations \citep{Bouchy2001,Eggenberger2004,Kervella2017,Thevenin2026}.
AI Phe provides a complementary eclipsing-binary benchmark, with accurately determined masses, radii, and effective temperatures for two components probing distinct evolutionary states with convective envelopes \citep{Maxted2020,Miller2020,Zhou2025}.
Together, these systems would provide demanding tests of whether the resulting one-dimensional structures can reproduce precise stellar constraints using the closure alone.

Beyond these solar-like benchmarks, extending the comparison to lower-gravity, higher-Mach-number convection, in particular in red-giant envelopes, would provide a valuable test of the closure in a substantially different regime.
More broadly, the present framework suggests a natural route towards stellar models in which radiative, convective, and transitional regions are treated as limiting cases of a single mean-field description.
Also, because the closure is formulated at the level of a phase-space distribution rather than a scalar mixing length, its extension beyond spherical symmetry would naturally involve vector and tensor moments of the velocity distribution.
This makes two- and three-dimensional mean-field models, including rotating stars with intrinsically anisotropic convective statistics, a particularly promising direction for future developments.

\begin{acknowledgements}
We thank the referee for thoughtful and constructive comments that helped clarify several aspects of the framework and improve the overall presentation of the manuscript.
We are grateful to the M3DIS team for making their radiation-hydrodynamics simulations publicly available.
Access to these data made it possible to confront the proposed closure with realistic three-dimensional stellar surface convection models and was therefore central to the present work.
This work was funded by the European Research Council (ERC) under the Horizon Europe programme (Synergy Grant agreement No. 101071505: 4D-STAR). 
While partially funded by the European Union, views and opinions expressed are however those of the authors only and do not necessarily reflect those of the European Union or the European Research Council. 
Neither the European Union nor the granting authority can be held responsible for them.
\end{acknowledgements}

\bibliographystyle{aa}
\bibliography{src}

\appendix
\nolinenumbers

\section{Relation between \(\{s,T\}\) and baroclinic vorticity}
\label{app:baroclinic_limit}

In this appendix we clarify under which local assumptions the phase-space divergence \(\{s,T\}\) may be related to baroclinic vorticity production.
The purpose is not to derive a general identity valid for arbitrary flows, but rather to make explicit the physical intuition underlying the remixed conversion law used in Sect.~\ref{sec:plume_fraction}.
The picture we have in mind is that of a plume descending through the upper convective layers while remaining locally more expanded than the surrounding bulk.
Such a configuration naturally creates a strong contrast in specific volume across the plume--bulk interface.
Combined with the ambient pressure stratification, this contrast generates baroclinic vorticity precisely in the region where remixed conversion between the two populations is expected to be most efficient.

In addition to the low-Mach and high-Reynolds assumptions used throughout the paper, we further assume:

\begin{enumerate}[(i)]
        \item that the relevant local kinematics may be described in the tangent plane to the isobar;
        \item that, in that plane, the antisymmetric part of the velocity gradient dominates the local deformation;
        \item that local displacements are quasi-adiabatic (\(\delta s \simeq 0\)), so that, for motions within the tangent plane to the isobar (\(\delta p = 0\)), the specific volume of the displaced element is preserved to leading order: \(\delta v \simeq 0\);
        \item that the local vorticity balance is itself dominated by baroclinic production.
\end{enumerate}

Under these assumptions, we show below that \(\{s,T\}\) may be interpreted as the logarithmic growth rate of the squared baroclinic vorticity amplitude in the isobaric plane.

\subsection{Projected local kinematics on the isobar}
\label{app:baroclinic_limit_kinematics}

We begin from
\begin{equation}
        \{s,T\} = \{v,p\} = -\grad_{u} v \cdot \grad_{r} p,
\end{equation}
where the last equality follows from \(\grad_{u} p=\vec 0\).
We now restrict attention to the local tangent plane to the isobar, since this is the natural plane along which pressure-preserving displacements occur, and also the one expected to contain the strongest contrasts in specific volume between plumes and bulk.

Assumption (ii) states that, in that plane, the antisymmetric part of the velocity gradient dominates the local kinematics.
Writing \(\delta\vec\xi_\parallel\) and \(\delta\vec u_\parallel\) for the position and velocity increments projected onto that plane, we then have
\begin{equation}
        \label{eq:local_kinematics_vorticity}
        \delta\vec u_\parallel
        \simeq
        \mathrm W \cdot\delta\vec \xi_\parallel
        \qquad \mathrm W = \frac12\left(\grad_{r}\vec u - (\grad_{r}\vec u)^\dag \right)_\parallel.
\end{equation}
In two dimensions, any antisymmetric operator is proportional to the planar rotation tensor.
We therefore introduce the scalar vorticity \(\omega\) in the isobar by writing
\begin{equation}
        \mathrm W
        =
        \frac{1}{2}\, \omega\mathrm J,
\end{equation}
where \(\mathrm J\) denotes the rotation tensor by \(+\pi/2\) in the tangent plane.
The local kinematics \eqref{eq:local_kinematics_vorticity} can then be inverted to yield 
\begin{equation}
        \label{eq:local_kinematics_vorticity_reversed}
        \delta\vec \xi_\parallel \simeq -\frac{2}{\omega}\,\mathrm J \cdot \delta\vec u_\parallel.
\end{equation}

By assumption (iii), such a displacement preserves the specific volume to leading order, so that
\begin{equation}
        \delta v
        =
        \grad_{r} v \cdot \delta\vec\xi_\parallel
        +
        \grad_{u} v \cdot \delta\vec u_\parallel
        \simeq 0.
\end{equation}
Substituting Eq.~\eqref{eq:local_kinematics_vorticity_reversed} into this relation then yields
\begin{equation}
        \grad_{u} v \simeq \frac{2}{\omega}\,\mathrm J \cdot \grad_{r} v.
\end{equation}
This relation expresses the fact that, when the local interfacial motion is predominantly rotational, the velocity-space sensitivity of \(v\) may be read as a rotated spatial gradient within the isobaric plane.

\subsection{From \(\{s,T\}\) to baroclinic vorticity growth}
\label{app:baroclinic_limit_vorticity}

Substituting the projected relation above into the bracket yields
\begin{equation}
        \{s,T\}
        =
        \{v,p\}
        \simeq
        -\frac{2}{\omega}\,
        \left(\mathrm J\cdot\grad_{r} v\right)\cdot\grad_{r} p.
\end{equation}
Since \(\mathrm J\) rotates vectors within the tangent plane to the isobar, this expression is equivalent to the baroclinic source projected onto that plane : 
\begin{equation}
        \{s,T\}
        \simeq
        -\frac{2}{\omega}\,
        \left(\grad_{r} v \times \grad_{r} p\right)_\parallel.
\end{equation}

Under assumption (iv), the local vorticity balance is itself dominated by this baroclinic term, so that, along the motion,
\begin{equation}
        D_t \omega \simeq -\left(\grad_{r} v \times \grad_{r} p\right)_\parallel.
\end{equation}
Combining the two relations then gives
\begin{equation}
        D_t \omega \simeq \dfrac{1}{2} \,  \{s,T\}\, \omega.
\end{equation}
Integrating along the trajectory finally yields
\begin{equation}
        \omega(t)
        =
        \omega_0
        \exp \left(
        \frac12 \int \{s,T\} \,dt
        \right).
\end{equation}
Under assumptions (i)--(iv), the phase-space divergence \(\{s,T\}\) may thus be read as the rate at which the baroclinic vorticity accumulates in the tangent plane during the descent.

\end{document}